\documentclass[letterpaper,useAMS,usenatbib]{mn2e}
\usepackage{graphicx}
\usepackage{float}
\usepackage{amssymb}
\usepackage{wasysym}
\usepackage{pifont}
\usepackage{bm}
\usepackage[Symbol]{upgreek}
\usepackage[usenames]{color}

\def\jref@jnl#1{{\rm#1}}

\def\aj{\jref@jnl{AJ}}                   
\def\araa{\jref@jnl{ARA\&A}}             
\def\apj{\jref@jnl{ApJ}}                 
\def\apjl{\jref@jnl{ApJ}}                
\def\apjs{\jref@jnl{ApJS}}               
\def\ao{\jref@jnl{Appl.~Opt.}}           
\def\apss{\jref@jnl{Ap\&SS}}             
\def\aap{\jref@jnl{A\&A}}                
\def\aapr{\jref@jnl{A\&A~Rev.}}          
\def\aaps{\jref@jnl{A\&AS}}              
\def\azh{\jref@jnl{AZh}}                 
\def\baas{\jref@jnl{BAAS}}               
\def\jrasc{\jref@jnl{JRASC}}             
\def\memras{\jref@jnl{MmRAS}}            
\def\mnras{\jref@jnl{MNRAS}}             
\def\pra{\jref@jnl{Phys.~Rev.~A}}        
\def\prb{\jref@jnl{Phys.~Rev.~B}}        
\def\prc{\jref@jnl{Phys.~Rev.~C}}        
\def\prd{\jref@jnl{Phys.~Rev.~D}}        
\def\pre{\jref@jnl{Phys.~Rev.~E}}        
\def\prl{\jref@jnl{Phys.~Rev.~Lett.}}    
\def\pasa{\jref@jnl{PASA}}               
\def\pasp{\jref@jnl{PASP}}               
\def\pasj{\jref@jnl{PASJ}}               
\def\qjras{\jref@jnl{QJRAS}}             
\def\skytel{\jref@jnl{S\&T}}             
\def\solphys{\jref@jnl{Sol.~Phys.}}      
\def\sovast{\jref@jnl{Soviet~Ast.}}      
\def\ssr{\jref@jnl{Space~Sci.~Rev.}}     
\def\zap{\jref@jnl{ZAp}}                 
\def\nat{\jref@jnl{Nature}}              
\def\iaucirc{\jref@jnl{IAU~Circ.}}       
\def\aplett{\jref@jnl{Astrophys.~Lett.}} 
\def\apspr{\jref@jnl{Astrophys.~Space~Phys.~Res.}}
\def\bain{\jref@jnl{Bull.~Astron.~Inst.~Netherlands}} 
\def\fcp{\jref@jnl{Fund.~Cosmic~Phys.}}  
\def\gca{\jref@jnl{Geochim.~Cosmochim.~Acta}}   
\def\grl{\jref@jnl{Geophys.~Res.~Lett.}} 
\def\jcap{\jref@jnl{JCAP}}      %
\def\jcp{\jref@jnl{J.~Chem.~Phys.}}      
\def\jgr{\jref@jnl{J.~Geophys.~Res.}}    
\def\jqsrt{\jref@jnl{J.~Quant.~Spec.~Radiat.~Transf.}}
\def\memsai{\jref@jnl{Mem.~Soc.~Astron.~Italiana}}
\def\nphysa{\jref@jnl{Nucl.~Phys.~A}}   
\def\physrep{\jref@jnl{Phys.~Rep.}}   
\def\physscr{\jref@jnl{Phys.~Scr}}   
\def\planss{\jref@jnl{Planet.~Space~Sci.}}   
\def\procspie{\jref@jnl{Proc.~SPIE}}   

\def\cm2og{\,\mathrm{cm}^2\,\mathrm{g}^{-1}}

\def\Lcal{\mathcal L}
\def\Mcal{\mathcal M}

\title[effect of cD galaxy on host cluster]{
Gravitational and distributed heating effects of a
cD~galaxy on the hydrodynamical structure of its host cluster
}
\author[Saxton \& Wu]{Curtis J. Saxton$^{1}$\thanks{E-mail:
cjs2@mssl.ucl.ac.uk (CJS); kw@mssl.ucl,ac,uk (KW)
} 
\& Kinwah Wu$^{1}$
\\
$^{1}$Mullard Space Science Laboratory, University College London,
Holmbury St Mary, Surrey RH5 6NT, UK
}

\begin{document}

\date{Accepted 2013 November 7. Received 2013 November 5; in original form 2012 December 27}

\pagerange{\pageref{firstpage}--\pageref{lastpage}} \pubyear{2013}

\maketitle

\label{firstpage}

\begin{abstract} 

\noindent 
We investigate the effects of a cD galaxy's gravity
   and AGN heating of the host galaxy cluster. 
We consider a standard prescription for the hydrodynamics,
   with the structures determined by
   mass continuity, momentum and energy conservation equations
   in spherical symmetry.
The cluster comprises a dark matter halo (DM)
   and ionized X-ray emitting intracluster gas (ICM),
   which jointly determine the gravitational potential.
The cD galaxy is an additive gravitational potential component.
The DM assumes a polytropic equation of state
   (determined by its microphysics),
   which could be non-radiative self-interacting particles
   or more exotically interacting particles.   
The AGN provides distributed heating,
   counteracting radiative cooling.
Stationary density and velocity dispersion profiles
   are obtained by numerically integrating the hydrodynamic equations
   with appropriate boundary conditions.    
The minimum gas temperature in the cluster core
   is higher when a cD galaxy is present than when it is absent.
The solutions also yield
    a point-like mass concentration exceeding a minimum mass:
    presumably the AGN's supermassive black hole (SMBH).
Consistency with observed SMBH masses
    constrains the possible DM equations of state.
The constraints are looser when a cD galaxy is present.    
Distributed (AGN) heating alters cluster global properties,
   and also reduces the lower limits for the central point-mass,
   for the preferred DM models in which the dark particles
   have greater heat capacity than point particles.
Eluding these constraints would require
   dominant non-spherical or anisotropic effects
   (e.g. bulk rotation, non-radial streaming,
   asymmetric lumps
   or a strong magnetic field).
\end{abstract}


%
%


%

\section{Introduction}

In galaxy clusters, most of the visible matter exists as
   the X-ray emitting gas of the intracluster medium (ICM),
   which is outweighed by the dark matter halo
   (DM) inferred to bind the system together.
It has long been recognized that gas cooling in undisturbed clusters
   must weaken central pressure support,
   leading to gas inflow from the outskirts.
In the conventional `cooling flow' models
\citep[e.g][]{fabian1977,cowie1977,stewart1984,nulsen1986,johnstone1992}
   runaway inflows would deposit multiphase cold gas throughout
   a 0.3~Mpc cool core
   at rates of $10^1$--$10^3~m_\odot~{\rm yr}^{-1}$.
Observationally, these cold condensates do not occur
   in the predicted amounts,
   and the ICM appears single-phase
\citep[e.g.][]{ikebe1997,boehringer2001,david2001,kaastra2001,
	tamura2001,peterson2001,
	molendi2001,matsushita2002,donahue2004,peterson2006}.
In the cool cores of clusters and groups,
   the gas temperatures are observed to drop,
   but rarely
   \citep[Centaurus:][]{sanders2008}
   more than a factor $\sim2$--5
   below the peak temperature
   \citep[e.g.][]{sakelliou2002,ettori2002,johnstone2002,peterson2003,
	voigt2004,xue2004,bauer2005,zhang2006,reiprich2009,
	osullivan2011,moretti2011,bulbul2012}.
In some cool cores, the temperature actually rises
   at small radii around a central galaxy
   \citep{osullivan2007a,sun2009}.

To remedy this runaway cooling-induced inflow problem,
   non-gravitational heating, usually by active galaxies (AGN),
   was invoked to staunch the inflows
   \citep[see e.g. reviews by][]{peterson2006,mcnamara2007}.
Yet it is questionable whether the heating can be sufficiently well distributed
   to attain finely balanced `feedback'
   that actively controls clusters of all types. 
Although bubbles blown by active galaxies
   should contain enough heat for clusters in general 
   \citep[e.g.][]{churazov2001,birzan2004,dunn2005},
   it is uncertain how well the hot plasma in the bubble would mix with the ICM
   \citep[e.g.][]{dursi2007}. 
It is also unknown whether the compression of the ICM
   by bubble-generated shocks 
   actually exacerbates or halts the cooling
   \citep[e.g.][]{brighenti2003,conroy2008}.

The nature of dark matter remains speculative and contentious.
On the one hand, $N$-body simulations
   reproduce large-scale cosmic structures resembling those observed;    
   on the other hand, simulations also produce results
   inconsistent with various observations.  
If the dark matter is assumed to experience only gravitation,
   the abundance of luminous substructure is overpredicted,
   for Milky Way satellites
   \citep[e.g.][]{klypin1999,moore1999},
   for galaxies in some groups
   \citep{donghia2004},
   in cosmic voids
   \citep{tikhonov2008,peebles2010}
   and at higher redshifts
	\citep{miller2013}.
More problematically,
   the largest observed satellite galaxies
   are less massive than the corresponding predicted subhaloes
   \citep{boylan2011,boylan2012}.
Collisionless DM models also predict singular central density {\em cusps}
   of dark matter in galaxies and clusters
   \citep{dubinski1991,nfw1996,navarro2004,merritt2005}.
To date, zero galaxies have unambiguously proven dark matter cusps.
Instead, the evidence from diverse galaxy types
   either strongly requires, favours or allows
   dark matter {\em cores} of nearly uniform density
\citep[e.g.][]{flores1994,moore1994,burkert1995,salucci2000,kelson2002,
	kleyna2003,goerdt2006,
	gentile2004,deblok2005,thomas2005,kuzio2006,
	gilmore2007,weijmans2008,oh2008,nagino2009,inoue2009,
	deblok2010,pu2010,murphy2011,memola2011,walker2011,
	jardel2012,amorisco2012,agnello2012}.
These imply that dark matter is not as simple as previously thought. 
It could well self-interact (i.e\ self-interacting dark matter, SIDM),
   with the soft central cores emerging universally due to dark pressure support
   \citep[e.g.][]{spergel2000,firmani2000,ahn2005,ackerman2009,
	loeb2011,vogelsberger2012,peter2012,rocha2012}.

An alternative remedy to rectify the cusp problem
   is the injection of mechanical energy into the halo.  
Through a `feedback' mechanism
   that invokes supernovae and stellar wind outflows, 
   the cusps are shaken flat
   \citep[e.g.][]{navarro1996,binney2001}.    
With ad~hoc supernova recipes,
   this seems to work in some simulations of gassy dwarf galaxies 
   \citep[e.g.][]{mashchenko2006,governato2010},
   but may be energetically impossible for DM-dominated dwarf spheroidals
   \citep{penarrubia2012,kimmel2013}.
A total blowout of the baryons is still insufficient to erase cusps 
   in a larger galactic disc \citep{gnedin2002}.
Nevertheless, 
  the use of black-box recipes for subgrid stellar physics 
  is computationally convenient, 
  as the assumption of collisionless dark matter is retained.  
\cite{sellwood2009}
   reviews other baryonic mechanisms
   speculated to destroy dark cusps at galaxy scales.
Similar energy-injection theories propose that
   the cusps of elliptical galaxies were erased by AGN feedback
   \citep{peirani2008a}. 

At cluster scales the observational evidence
   about the nature of dark matter
   is less settled than for galaxies.
At least some
   X-ray, kinematic and gravitational lensing studies of clusters
   require or allow DM cores
   \citep{sand2002,sand2008,ettori2002,halkola2006,halkola2008,
	voigt2006,rzepecki2007,newman2009,zitrin2009,richtler2011},
   while others allow cuspy models
   \citep{athreya2002,buote2004,pointecouteau2005,gavazzi2005,diego2005,
	vikhlinin2006,saha2008,richard2009}.
Inclusion of information from the inner 100 or 30\,kpc radii
   (especially stellar kinematics)
   tends to favour a core rather than cusp
   \citep[e.g.][]{gavazzi2005,sand2008}.
It is sometimes claimed that
   asymmetric clusters are mergers
   (capable of disproving DM collisionality)
   but the evidence revolves around assumed motions and projected geometries
   of a handful of special objects.
A particular `Bullet Cluster',
   studied in X-rays and gravitational lensing,
   has been interpreted as a head-on merger
   with interpenetrating, non-interacting haloes
   \citep[{1E0657$-$56:}][]{clowe2006,randall2008},
   but since ICM shocks have not affected star formation as expected
   \citep{chung2009}
   a different story may be necessary.
Another merger taken at face value
   ({Abell 520})
implies that dark matter behaves like gas,
concentrated and separated from the collisionless galaxies
\citep{mahdavi2007,jee2012}.
\cite{williams2011} have investigated a gravitational lensing cluster
   ({Abell 3827})
   in which the haloes of the innermost elliptical galaxies
   appear displaced from the stars,
   perhaps due to drag forces in the cluster halo.

The feedback recipes for eliminating cusps and staunching cooling flows
   could be elaborated for decades indecisively.
It is therefore worthwhile to study alternative theories
   that produce the required structures inexorably.
Our cluster model
   \citep{saxton2008}
   revisits the cooling-induced inflow scenario
   in a more complete and consistent implementation,
   to reassess the natural behaviour of gas in a quiescent halo.
We consider versions of halo physics (including SIDM)
   which produce DM cores naturally.
Given a sensible radius and total cluster mass,
   we infer constraints on the DM parameters
   and the central object of the cluster.
A favoured domain of DM thermal microphysics
   yields cores of realistic size
   ($\sim10^1$--$10^2$kpc).
All stationary solutions of our model
   have non-zero gas temperatures
   and possess a central point-mass exceeding some minimum.
Requiring consistency with the masses of observed black holes
   \citep[e.g.][]{houghton2006,inada2008,cappellari2009,
		gebhardt2011,mcconnell2011,mcconnell2012}
   implies joint constraints on the dark matter physics
   and gas inflow rate.
The tightest constraints on the continuity of the gas inflow occur at kpc radii,
   suggesting that this is the natural site
   for cold gas dropout (and star formation)
   during external disturbances.
For the favoured DM models,
   it was also found that an inner portion of the dark halo
   teeters on the brink of gravitational collapse.
The dark mass involved is consistent with observed supermassive black holes
   (SMBH),
   hinting that these objects
   could feed non-radiatively in `dark gulping' events.   

This modelling omitted the stellar mass distribution
   of the cD galaxy that should realistically
   reside at the centre of an inflow of cooling gas,
   and surrounding the black hole.
Stellar density is significant compared to dark matter
   within the half-light radii of elliptical galaxies
   \citep[e.g.][]{loewenstein1999,kronawitter2000,
	ferreras2005,thomas2007,bolton2008,saxton2010,norris2012}.
In this work, we introduce such a galaxy,
   and determine how it modifies the structure of
   the cluster's gas inflow and dark halo. 
In particular, we investigate how the presence of a cD galaxy
   would affect the temperature floor of the ICM gas. 

We also consider how AGN heating could affect cluster properties.
To be effective,
   AGN power needs to be isotropically deposited in the cool core.
Jet effects are directional \citep{vernaleo2006}
   but they may be a feasible heating process
   if the jet axis slews and realigns
   widely enough between active episodes
   \citep{babul2012}.
These conditions are conceivable:
   some observed pairs of giant radio lobes suggest
   large angular slews between outbursts $\la100$~Myr apart
   \citep[e.g.][]{dunn2006},
   while a blazar-like tidal disruption jet
   may precess and nutate through smaller angles on weekly time-scales
   \citep[e.g. Sw~J1644+57;][]{saxton2012}.
Streaming cosmic rays could provide a more innately isotropic
   heating process than jets
   \citep[e.g.][]{fujita2011}.
In this paper
   we optimistically {\em assume} isotropic heating
   and consider various forms of radially distributed AGN power.
We compare cases where the cD galaxy is active or inactive.

The paper is organized as follows:
 Section~\ref{s.structure}
   describes the hydrodynamical formulation
   and the ingredients of the cluster model; 
 in Section~\ref{s.results},
   we show the results of our calculations;
  Section~\ref{s.discussion}
   is a discussion;
 and \S\ref{s.conclusions} presents a brief summary of our findings.

\section{Hydrodynamical structure}    
\label{s.structure}

\subsection{Constituents}
\label{s.constituents}

We consider spherical galaxy clusters 
   consisting of  two free, coterminous mass components
   (subscripted $i=1, 2$), 
   interacting only via their shared gravitational potential
   \citep[following][]{saxton2008}.
The first component is the intracluster medium (ICM, $i=1$), 
   which is hot ionized X-ray emitting gas.
The most massive component is non-radiative dark matter ($i=2$).  
The particles in the two components are classical, 
   and their velocity distributions are isotropic. 
Their effective degrees of freedom, $F_i$,  
   are determined by the corresponding microphysics.
The internal energy density is
   $\epsilon_i=F_iP_i/2$,
   where $P_i$ is the partial pressure.

Each component has the generic equation of state,%
\begin{equation}
	P_i=\rho_i\sigma_i^2=s_i\,\rho_i^{\gamma_i} \ , 
\label{eq.state}
\end{equation}
   where
   $\rho_i$ the density,
   $\sigma_i$ is a velocity dispersion and 
   $s_i$ is the pseudo-entropy.  
Adiabatic processes leave $s_i$ constant;
   objects with uniform $s_i$ are `polytropes.'
The adiabatic index is given by   
\begin{equation} 
   \gamma_i =    1+ \frac{2}{F_i} \  .     
\end{equation}
Many physical scenarios entail a condition such as equation (\ref{eq.state}).
For an ideal gas, the index is the ratio of specific heats,
    $\gamma=c_{_P}/c_{_V}$.
For a monatomic gas or a fully ionized plasma, $F=3$ and $\gamma ={\frac53}$; 
   for a relativistic or radiation-dominated gas, $F=6$ and $\gamma={\frac43}$. 
Composite particles can have higher $F$
   because of their rotations and other freedoms:
   e.g. for a gas of mass dipoles, $F=5$ and $\gamma={\frac75}$.
Polyatomic gases or SIDM `dark molecules'
   \citep[e.g.][]{alves2010,kaplan2010a}
   might have even higher $F$.
An isothermal gas has infinite heat capacity,
   corresponding to $F\rightarrow\infty$ and $\gamma =1$. 
The classic \cite{plummer1911} model requires $F=10$ and $\gamma={\frac65}$.

Some boson-condensate and scalar field dark matter models give
   $F=2$ and $\gamma=2$
   \citep{arbey2003,boehmer2007,harko2011c,chavanis2011}
   though other values are possible \citep{peebles2000}.
These SIDM need not be seen as consisting of
   distinctly localized particles
   (collisional or otherwise).
Some theories propose
   phase changes in the outskirts of haloes
   \citep{arbey2006,slepian2012}
   which we need not consider here.
   
Theories of thermostatistics for systems with long-range interactions
   \citep[e.g.][]{tsallis1988}
   predict that equilibria of {\em collisionless} spheres are polytropes,
   probably with non-integer $F$
   \citep{plastino1993,nunez2006,zavala2006,vignat2011}.
Two collisionless dark matter species
   can act together as a single anisotropic dark fluid
	\citep[see][]{harko2011b,harko2012}, 
   but this would acquire an equation of state more complicated than
	equation (\ref{eq.state}).

For a single adiabatic fluid with non-singular central density
   and constant $s$,
   the density profile is a classical polytropic sphere
   \citep{lane1870,emden1907,chandrasekhar1939,viala1974,viala1974b}.
If $-2<F<10$ then the density truncates at a finite outer radius ($R$),
   whereas the outskirts of collisionless haloes blur away as $\rho\sim r^{-3}$.
The self-truncation of a polytropic halo seems more consistent with
   the steeper outer profiles observed in some clusters
   \citep{nevalainen1999,broadhurst2005a,diego2005,umetsu2008}
   and galaxies
   \citep[e.g.][]{kirihara2013}.
A core of nearly uniform density
    fills a larger part of the sphere if $F$ is smaller.
The scale radius $R_2$
   where the density slope
   $\mathrm{d}\ln\rho/\mathrm{d}\ln r=-2$
   occurs outside the core near $0.520R$, $0.101R$ and $0.0459R$
   if $F=3, 8$ and $9$, respectively.
The core boundary radius $R_1$ (slope $-1$)
   occurs near $0.379R$, $0.0636R$ and $0.0285R$, respectively
   \citep[see also table~A1 of][]{saxton2013}.
Including stars or another gravitating fluid
   alters these proportions slightly.

Our implicit assumption of locally isotropic particle velocities
   is justified in a variety of DM theories:
   if the halo has previously been well mixed by violent relaxation
   and shaking of the potential;
   if SIDM consists of collisional particles with short mean-free-path;
   if the SIDM is supported by dark forces,
   resembling the plasma effects that mediate collisionless shocks;
   or if $P$ and $\rho$ are aspects of a smooth non-classical field
   (and $\sigma^2=P/\rho$ is merely a derived quantity).
Isotropy is necessarily implied in the central pressure-supported core.
If the outermost matter is also isotropic,
   then the halo can truncate at an outer radius, $R$.
If the outskirts are too collisionless,
   then they may blur into the cosmic background.
In that case, our finite models serve as an idealized setting,
   and attention should focus on
   the core where gas cooling and SIDM effects are strongest.

Many authors interpret SIDM as
   point-like ($F_2=3$) self-scattering particles
   (which is convenient to implement in $N$-body simulations).
Some simulations predict oversized cluster cores,
   prompting suggestions that scattering cross-sections are weak
   \citep[$\varsigma<1\cm2og$;][]{yoshida2000b,dahle2003,arabadjis2002,
	katgert2004,vogelsberger2012,rocha2012}.
This limitation is
   not the only conceptually simple possibility.
Alternatively, we would vary the heat capacity:
   the domain $7\la F_2<10$ provides realistic core sizes
   ($R_1<0.16R$)
   without restriction on $\varsigma$
   \citep{saxton2008,saxton2010,saxton2013}.

\subsection{Formulation}
\label{s.formulation}

The gravitational potential ($\Phi$)  
   and field ($\bm{g}= -\nabla\Phi$)
   are derived from the Poisson equation,
   for the densities present ($\rho_i$),
\begin{equation}
	\nabla^2\Phi = 4\uppi G\sum_i\rho_i 	\ .
\end{equation}
Because each fluid density will be calculated
   simultaneously with the potential,
   our formulation implicitly includes
   the effect of adiabatic contraction of the dark halo, 
   which is driven by the cooling-induced baryonic inflow
   \citep{blumenthal1986}.

The mass, momentum and energy conservation equations of the system are   
\begin{equation}
	{\upartial\over{\upartial t}}\rho_i
	+\nabla\cdot\rho_i\bm{v}_i = 0
	\ ,
\end{equation}
\begin{equation}
	{\upartial\over{\upartial t}}\rho_i{\bm{v}}_i
	+\nabla\cdot\rho_i{\bm{v}}_i {\bm{v}}_i
	+\nabla\rho_i\sigma_i^2
	=
	\rho_i\,{\bm{g}}
	\ ,
\end{equation}
\begin{equation}
	{\upartial\over{\upartial t}}\epsilon_i
	+\nabla\cdot\left({
		\epsilon_i+\rho_i\sigma_i^2
	}\right){\bm{v}}_i
	=
	\rho_i{\bm{v}}_i\cdot{\bm{g}} + {\mathcal L}_i \ .  
\end{equation} 
Here ${\bm{v}}_i$ is the flow velocity.
The source function ${\mathcal L}_i$ in the energy equation 
   takes account of all non-gravitational heating and cooling processes. 
We note that conventional `cooling flow' models
   do not generally include the kinetic terms. 
Our formulation treats the kinetic terms explicitly, 
  and all mass components participate in the gravitational interaction.  
  
We consider stationary spherical galaxy clusters. 
Therefore,  $\upartial/\upartial t\rightarrow 0$,
   $\upartial/\upartial\theta = 0$
   and $\upartial/\upartial\phi =  0$.  
In stationary solutions, the inflow or outflow of each component
   ($\dot{m}_i\equiv 4\uppi r^2\rho_i v_i$)
   is spatially constant.
We assume that the dark halo is in dynamical equilibrium,
   with $v_2=0$ and $\dot{m}_2=0$ everywhere.
The gas however flows inwards gradually, due to radiative cooling
   ($\dot{m}=\dot{m}_1<0$).
This forms a single-phase inflow. 
For the solutions relevant to physical galaxy clusters, 
   the inflow is subsonic everywhere 
   ($0<v_1^2<\gamma_1\sigma_1^2$).

We introduce a set of new fluid variables
   in terms of powers of the radial coordinate 
   for the hydrodynamic formulation:
\begin{eqnarray}
	\beta_{\rho_i}&\equiv&\rho_i\ r^{F_i/2}
		= \dot{m}_i / 4\uppi\beta_{v_i}
	\\
	\beta_{v_i}&\equiv& v_i\ r^{(4-F_i)/2}
	\\
	\beta_{\sigma_i}&\equiv& \sigma_i^2\ r  \ . 
\end{eqnarray}
Expressing the hydrodynamic variables in these power laws
   softens their behaviour 
   within the Bondi-like accretion spike that is inevitable 
   when a point gravitating mass exists at the origin
   \citep{bondi1952}.
Given values for the total mass at the outer boundary of the cluster,
   the mass of the gas ($i=1$) or dark matter ($i=2$)
   within any radius $r$
   can be obtained by numerical integration of an ODE,
\begin{equation}
	{{{\mathrm d}m_i}\over{{\mathrm d}l}}
	= 4 \uppi r^3 \rho_i
	= 4 \uppi \beta_{\rho_i}\ r^{(6-F_i)/2}
	\ , 
\label{eq.beta.mass}
\end{equation} 
   where $l\equiv \ln r$ is a log-radial coordinate.
The form of equation (\ref{eq.beta.mass})
   reveals that the mass profile $m_i(r)$
   naturally has steep inner gradients
   if $F_i>6$.
In the following equations,
   we abbreviate the {\em total} enclosed mass profile
   of gas plus DM plus stars as
\begin{equation} 
   m(r) = m_1(r) + m_2(r) + m_\bigstar(r) \ ,   
\label{eq,total.mass.star}
\end{equation} 
   (see \S\ref{s.galaxy} for details of $m_\bigstar$).

The Poisson and continuity equations can decouple
  into a set of first-order differential equations: 
\begin{eqnarray}
	{{\mathrm{d}\beta_{v_i}}\over{\mathrm{d}l}} 
	&\hspace{-3mm}=&\hspace{-3mm}
 	 \beta_{v_i}
	\left\{{
	   {{4-F_i}\over{2}}  
		-{{
		2\gamma_i \beta_{\sigma_i}- G m + {2\over{F_i}}H\beta_L r^c
	  }\over{\gamma_i\beta_{\sigma_i}(1-\Mcal^2) }}
	}\right\}\ ,
\label{eq.beta.v}
\end{eqnarray}
\begin{eqnarray}
	{{\mathrm{d}\beta_{\sigma_i}}\over{\mathrm{d}l}}
	&\hspace{-3mm}=&\hspace{-3mm}
	\beta_{\sigma_i} + 
     	{{2\left[{
		2\gamma_i\beta_{\sigma_i} \Mcal^2  
		-G m	 -(1-\gamma_i \Mcal^2)H\beta_L r^c
	}\right]
		}\over{	F_i\gamma_i(1-\Mcal^2)}}
	\ ,
\label{eq.beta.sigma}
\end{eqnarray}
\begin{eqnarray}
	{{\mathrm{d}\beta_{\rho_i}}\over{\mathrm{d}l}}
	&\hspace{-3mm}=&\hspace{-3mm}
	\beta_{\rho_i}  
	\bigg\{ 	{F_i\over{2}}	+
		{{
		2\gamma_i\beta_{\sigma_i} \Mcal^2
		-G m
		+{2\over{F_i}}H\beta_L r^c\ 
		}\over{\gamma_i\beta_{\sigma_i} (1-\Mcal^2)}}
	   \bigg\}     \ .
\end{eqnarray}
\begin{equation}
	{{\mathrm{d}s_i}\over{\mathrm{d}l}}
	=
	- {{2BH\beta_{\rho_i}r^c
		}\over{F_i\,\beta_{v_i} \sqrt{\beta_{\sigma_i}}}}\ s_i
	\  . 
\end{equation}  
The inflow Mach number is given by    
\begin{eqnarray}  
  \Mcal_i^2 & = & \frac{\beta_{v_i}^2 r^{F_i-3}}{\gamma_i \beta_{\sigma_i}} \ ,
\end{eqnarray}  
 and its profile is    
\begin{eqnarray}
	{{\mathrm{d}\Mcal_i^2}\over{\mathrm{d}l}}
	&=&
	{{\Mcal_i^2}\over{1-\Mcal_i^2}}
	\biggl[
	-4\left({{\Mcal_i^2+F_i}\over{F_i}}\right)
		+{{2(F_i+1)Gm}\over{(F_i+2)\beta_{\sigma_i} }}
	\nonumber\\
	&&\hspace{2cm}
	-(1+\gamma_i \Mcal_i^2)
	{{
		2H\beta_L\ r^c
	}\over{
		(F_i+2)\beta_{\sigma_i}
	}}
	\biggr]
	\  .   
\label{eq.dMMdl}
\end{eqnarray}
Mathematically, there is some redundancy among these ODEs.
Practically, it is advantageous to integrate all of them simultaneously,
   because the numerical step-size control
   is inhibited and becomes more cautious
   at certain difficult parts of the radial profile,
   which prevents overstepping into unphysical conditions
   (e.g. where $\Mcal_i>1$, $\Mcal_i<0$ or $s_i<0$).

\subsection{Gas inflow, cooling and heating}

The terms $\beta_L$, $c$ and $H$ in equations
   (\ref{eq.beta.v})--(\ref{eq.dMMdl}) 
   describe the cooling and heating processes.
The ICM in massive clusters is highly ionized, implying that $F_1=3$.
The hot gas is cooled primarily by emitting thermal free--free X-rays. 
The thermal free--free cooling rate is
   $\propto \rho_{1}^2\sigma_{1}$ \citep[see][]{rybicki}, 
   with $\sigma_1=\sqrt{kT/\mu m_\mathrm{u}}$
   being the thermal velocity dispersion of gas.
We introduce a source term for the central AGN,
   which provides radiative and/or mechanical heating.   
We parametrize the AGN heating term as a power law of the radial distance $r$. 
The combined heating/cooling function then takes the form:    
\begin{equation}
	{\mathcal L}_1=A\,r^{-\nu} -B\rho_{1}^2\sigma_{1}
	\ . 
\label{eq.source}
\end{equation}
The parameter $B$ is a constant depending weakly on the plasma composition
   \citep[see][]{rybicki,saxton2008}.
The parameter $A$ specifies the heating rate, 
  and the index $\nu$ determines how heating is distributed spatially. 
Idealized radiative heating may be represented by $\nu=2$; 
   and centrally absorbed radiative heating by $\nu=2.5$. 
Heating via an efficient dispersive mechanical outflow
   would give $\nu\approx 0$ (i.e.\ uniform heating).  
An intermediate value would mimic  
   the shock heating and/or mixing of ascending radio lobes from the central AGN 
   \citep[e.g.][]{brueggen2001,churazov2001,saxton2001b,fujita2005a}.
Hybrid mechanical and radiative heating is parametrized by $0<\nu<2$, 
   and we take $\nu=1$ without losing generality.   
We will consider models with $\nu$ = 0, 1, 2 and $2.5$
   to investigate the effects of energy distribution
   by various types of AGN heating.
The expression for the heating/cooling function in equation (\ref{eq.source})
   implies that 
\begin{equation} 
 c=\frac{7}{2}-F_1 \  , 
\label{eq.cooling.1}
\end{equation} 
\begin{equation}
	\beta_L
	=  {{B\beta_{\rho_1} \sqrt{\beta_{\sigma_1}}}\over{\beta_{v_1}}}  \ ,   
\label{eq.cooling.2}
\end{equation}  
and the dimensionless heating ratio
\begin{equation}
	H
	\equiv{{B\rho_{1}^2\sigma_{1}-A\,r^{-\nu}}\over{B\rho_{1}^2\sigma_{1}}}
	=1-{{A\,r^{(2F_{1}-2\nu+1)/2}
		}\over{B\,s_{1}^{1/2}\beta_{\rho_{1}}^{(2F_{1}+1)/F_{1}} } }
	\ .
\label{eq.cooling.3}
\end{equation}

\subsection{Dark matter halo}

The dark matter is non-radiative;
   we set its cooling and heating functions to vanish
   ($A=B=H=\beta_L=c=0$).
We assume that dark matter distribution is essentially adiabatic.
The quasi-entropy, $s_2$,
   is uniform throughout the DM.
This may occur
   if the DM was well mixed in the cluster's prehistory,
   and $s_2$ measures the cumulative battering due to mergers.
Alternatively, $s_2$ might be a universal constant
   of the fundamental particle or dark field
	\citep[e.g.][]{peebles2000}.
Under the hydrostatic condition,
   the halo has $v_2=0$, ${\mathcal M}_2=0$ and $\beta_{v_2}=0$.
Then, we can simplify the hydrodynamic equations
   for $i=2$,
   leaving equation (\ref{eq.beta.mass})
   and one other gradient equation
\begin{equation}
	{{{\mathrm d}\beta_{\sigma_2}}\over{{\mathrm d}l}}
	= \beta_{\sigma_2} - {{2Gm}\over{F_2+2}}
	\ .
\label{eq.beta.sigma2}
\end{equation}
A version of the equation of state (\ref{eq.state})
   completes the description of halo structure:
\begin{equation}
	\beta_{\rho_2} = \left({
		\beta_{\sigma_2} / s_2
	}\right)^{F_2/2}
	\ .
\label{eq.beta.eos}
\end{equation} 
Additional ODEs can be written for $\beta_{\rho_2}$ and related quantities.
These are mathematically redundant,
   but harmless to numerical integration routines.
In this paper, we let $F_2$ be a parameter
   (in the domain $2\le F_2<10$).
We shall investigate how the cluster properties depend on its value,
  and hence set some constraints on the DM microphysics.

\subsection{cD galaxy} 
\label{s.galaxy}

We insert a central galaxy in some of the clusters.  
The cD galaxy has a \cite{sersic1968} light profile
   in the approximate deprojection by \cite{prugniel1997} 
   \citep[cf.\ the gasless galaxy model in][]{saxton2010,saxton2013}.
The stellar density profile is given by 
\begin{equation}
	\rho_\bigstar(r) = \rho_{\rm e}\  \left(\frac{r}{R_{\rm e}} \right)^{-p}
	\mathrm{e}^{ -b\left[{ \left(r/R_{\rm e} \right)^{1/n}-1}\right] }   \ , 
\end{equation}  
   where $\rho_{\rm e}$ is the stellar density
   at the half-light radius $R_{\rm e}$.
The indices $b$ and $p$ in the stellar density profile depend 
   on the shape index $n$ of the galaxy 
   \citep[see also][]{cotti1999,limaneto1999,marquez2000}.

The stellar mass enclosed within a radius $r$ is
\begin{equation}
	m_\bigstar(r) = 4\uppi n b^{n(p-3)} \mathrm{e}^b\ 
	\rho_{\rm e}R_\mathrm{e}^3\ 
	\Gamma\left[{
		n(3-p), b(r/R_{\rm e})^{1/n}
	}\right]	\ ,
\label{eq.mass.stars}
\end{equation}
   where $\Gamma(a,z)$ is the lower incomplete gamma function.    
In the present modelling,
   we adopt a total stellar mass
   $M_\bigstar\equiv{m}_\bigstar(\infty)=1.8\times10^{11}~m_\odot$
   and effective radius $R_{\rm e}=7.4$~kpc
   as fiducial values for the cD galaxy.
(In the natural units of \S\ref{s.numerical},
   we set $R_\mathrm{e}=0.03U_x$ and $M_\bigstar=0.02U_m$.)
The shape index is $n=4$,
   corresponding to the empirical de~Vaucouleurs profile
   \citep{devaucouleurs1948,devaucouleurs1953}.
The stellar mass profile is held fixed in the present calculations.
The dark matter and intracluster gas profiles
   are solved consistently in the presence of
   stellar mass profiles of the cD galaxy.

\subsection{Scaling and numerical integration}
\label{s.numerical}

The dimensional constants
   in this system of equations are 
   the gravitational constant $G$
   and the coefficient in the thermal free--free cooling function $B$.
Although $B$ depends slightly on the assumed metallicity, 
  we may roughly set a natural unit of length: $U_x\equiv B/G\approx0.246$~Mpc.
The velocity dispersion unit $U_\sigma\equiv1$
   is chosen corresponding to
   the isothermal velocity dispersion of a gas at temperature $kT=1$~keV. 
The implied unit of cluster-scaled luminosity (or power) 
   is then $U_L\approx1.44\times10^{45}~{\rm erg}~{\rm s}^{-1}$;
   the mass unit is $U_m\approx8.91\times10^{12}~m_\odot$;
   and the density unit is
   $U_\rho\approx6.01\times10^{14}~m_\odot~\mathrm{Mpc}^{-3}$.
Some cluster models are rescalable into homologously equivalent models
   as long as
   variables such as mass, temperature and velocity dispersion  
   are held in fixed ratios
   \citep[appendices~A and B,][]{saxton2008}.
The spatial measurements and the Mach numbers
   are unchanged under this rescaling scheme.
The composite quantities
   $\dot{m}^{3/2}/m(R)$ and $T_R/m(R)$
   are also invariants.
For numerical convenience,
   we set the units $B=G=1$ in our calculations. 

The hydrodynamic equations for the dark matter and intracluster gas 
  are integrated radially inwards towards the origin (the cluster centre). 
The minimum sufficient set of equations consists of
   equations (\ref{eq.beta.mass})--(\ref{eq.dMMdl}) for the gas,
   (\ref{eq.beta.mass}) and (\ref{eq.beta.sigma2}) for the halo,
   along with the algebraic expressions
   (\ref{eq.beta.eos}),
   (\ref{eq.mass.stars}) and
   (\ref{eq.cooling.1})--(\ref{eq.cooling.3}).
By default, in ordinary regions we express the ODEs of variables $y$
   in terms of the log-radial derivatives ($\mathrm{d}y/\mathrm{d}l$).
In regions with steeper spatial gradients,
   the code switches to an equivalent set of ODEs
   in another independent variable that provides shallower derivatives locally.
Near the halo's outer surface we use ODEs of the form
	$\mathrm{d}y/\mathrm{d}\beta_{\sigma_2}$.
Wherever the inflow approaches sound speed, we use
	$\mathrm{d}y/\mathrm{d}{\mathcal M}^2$.
In any circumstance,
   we employ the embedded eighth-order Runge--Kutta Prince--Dormand method
   with ninth-order error estimate
	\citep{prince1981,hairer2008}
   for the differential equation solver,\footnote{%
	We use mathematical routines from the {\sc Gnu Scientific Library}
	({\tt http://www.gnu.org/software/gsl/}).
	}
  as standard lower order Runge--Kutta
  and Bulirsch--Stoer methods are problematic 
  due to the stiffness of the hydrodynamic equations
  (at some locations).
The speed and accuracy of this method
   enables us to survey parameter domains comprehensively,
   to search for extrema in the output quantities,
   rather than picking on arbitrary example profiles.

\subsection{Boundary conditions and system parameters}
  
We consider the following outer boundary conditions: 
(i) the outer radius of the dark halo, $R$;
(ii) the total mass of the cluster,
    which includes the dark matter, the intracluster gas and cD galaxy, $m(R)$;
(iii) the gas inflow rate, $\dot{m}$; and
(iv) the outer gas temperature, $T_{R}$,
     which gives the corresponding gas Mach number, ${\mathcal M}_R$.
We impose that $0\le|{\mathcal M}_R|\le1$ 
  and consider a spatially uniform DM quasi-entropy $s_2$. 

We set the fiducial total cluster mass
   $m(R)=40U_m\approx3.57\times10^{14}~m_\odot$.
If our models were hydrostatic,
   we might approximately assume
   $\sigma_2\propto\sigma_1$ in the fringes
   and infer the other gas variables
   \citep{frederiksen2009}.
This is precluded since
   the inflow ($\dot{m}\neq0$) requires $\rho_1>0$ and $\sigma_1>0$
   at $r=R$ (where the DM truncates, $\sigma_2=0$).
We pick outer boundary conditions on the gas
   from a cosmologically plausible domain.
We set an outer gas temperature of $kT_R=1$~keV,
   representing likely conditions
   of shock-heated gas accreting from the cosmic background.
This is a reasonable extrapolation of 
   observed X-ray emission and Sunyaev--Zel'dovich effects,
   which show ICM temperatures declining to $\sim2$~keV around $r\sim2$~Mpc
   \citep[e.g.][]{
	simonescu2011,
	urban2011,
	eckert2013a,
	bonamente2013,
	ichikawa2013%
	}.

The gas inflow rate is $\dot{m}=10~m_\odot~{\rm yr}^{-1}$,
   which must be continuous from the outskirts inwards to the cool core
   and on to the central accretor.
This rate is modest compared to the `cooling flows'
   inferred for massive clusters,
   but great enough to overwhelm the $\sim0.1~m_\odot~\mathrm{yr}^{-1}$
   effects of a cD galaxy's stellar winds and supernovae
   \cite[e.g.][]{white1984,loewenstein1987a,sarazin1987,
	vedder1988,sarazin1989,mathews2003}.
We seek solutions in which the baryon fraction within $<R$
   matches a predefined cosmic value, 
   including the cosmic mean ($\approx0.16$).
Fortunately, these realistic models obtain outer gas densities ($\rho_1$)
   comparable to the cosmic mean.
The subtle effects of varying $T_R$ and $\dot{m}$
   were shown in \cite{saxton2008}.

By integrating radially inwards
   from physically justified external and global properties,
   we avoid the sensitivity to boundary conditions
   suffered by some other cooling flow models
   \citep[e.g. in isolated elliptical galaxies;][]{vedder1988}.
Integrating out from (unlucky) ad hoc central conditions
   can lead to unphysically high temperatures in the outskirts.
In a realistic universe, however,
   the cosmic background gas is indifferent
   to the properties of a cluster until it accretes.
It is the global mass and size of a cluster
   that constrain its internal details,
   not the details that control the bulk.
It might be interesting to explore these sensitivities,
   but in this work we implement
   the most robust method: computing from the outside inwards.

Once the integrator arrives at the origin,
   we test the consistency of conditions there.
The integration from $r=R$
   to the origin determines the mass profiles
   of the DM and the intracluster gas 
   and hence the central values
   $m_{1}(0)$ and $m_{2}(0)$. 
We may define 
\begin{equation}  
 m_{*} \equiv m(0) = m_{1}(0)+m_{2}(0)+m_{\bigstar}(0) \ ,  
\end{equation}  
  which is essentially the residual mass at the cluster centre.  
The \cite{prugniel1997} stellar profile
    does not contribute any mass at the origin, i.e.\ $m_{\bigstar}(0)=0$.
The residual mass $m_{*}$
   implies a point-like mass concentration at the origin,
   and we may interpret it as a massive black hole
   at the centre of the cluster.
We stress that $m_*$ is not an input parameter;
   it is a non-trivial output of each model calculation.
After integration, both the inner and outer boundary conditions are known,
   and we retrospectively find the total gas and DM masses:
   $M_1=m_1(R)-m_1(0)$ and $M_2=m_2(R)-m_2(0)$.
The baryon fraction follows directly.

\subsection{Allowed regions for physical solutions in the parameter space}
     
In the absence of the cD galaxy, 
  the integration of the hydrodynamic equations of the clusters 
  gives four types of solutions,  
  corresponding to four distinct regions in the parameter space.  
In \cite{saxton2008}, they were identified as follows.
\begin{enumerate}
\item
   the `{\it too cold}' zone:
   overcooling occurs at some finite radius.
     A zero-temperature shell falls freely inwards.
     This is inconsistent with the stationary assumption in the model.
\item
   the `{\it too fast}' zone:
   a supersonic break occurs at some radius.  
      This forbids a two-way communication between the interior and exterior,
         giving inconsistent mass fluxes.
\item
   the `{\it levity}' zone:
   the pressure support in the cluster is insufficient, 
      leading to an unphysical negative central mass.
\item
   the `{\it deep}' zone:
   the balance between radiative cooling of the gas
   and accretion warming everywhere 
      ensures that cold and fast inflow catastrophes
      do not occur in the cluster.
\end{enumerate}

Solutions corresponding to (i), (ii) and (iii) are unphysical. 
Stationary solutions corresponding to (iv) subject to the boundary conditions
   are acceptable physical solutions.  
In the presence of a cD galaxy, 
  the four types of solutions
  and their corresponding distinguishable zones
  in the parameters are also identified. 
However, 
   the zone borders for the cases with and without a central cD galaxy
   differ in detail.

\section{Results}
\label{s.results}

\subsection{Global properties and central mass}

\begin{figure}
\centering
$\begin{array}{cc}
\includegraphics[width=82mm]{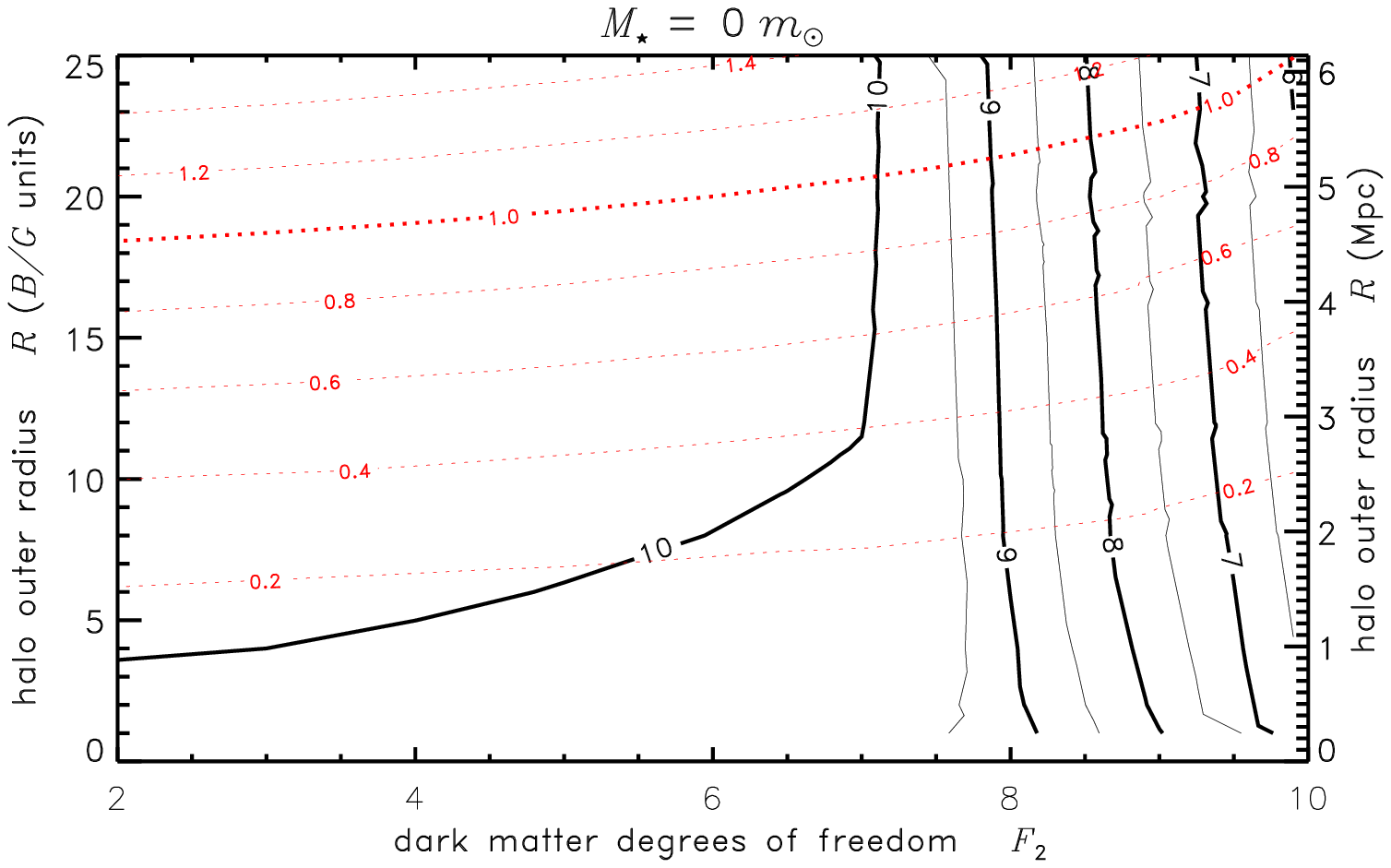}
\\
\\
\\
\includegraphics[width=82mm]{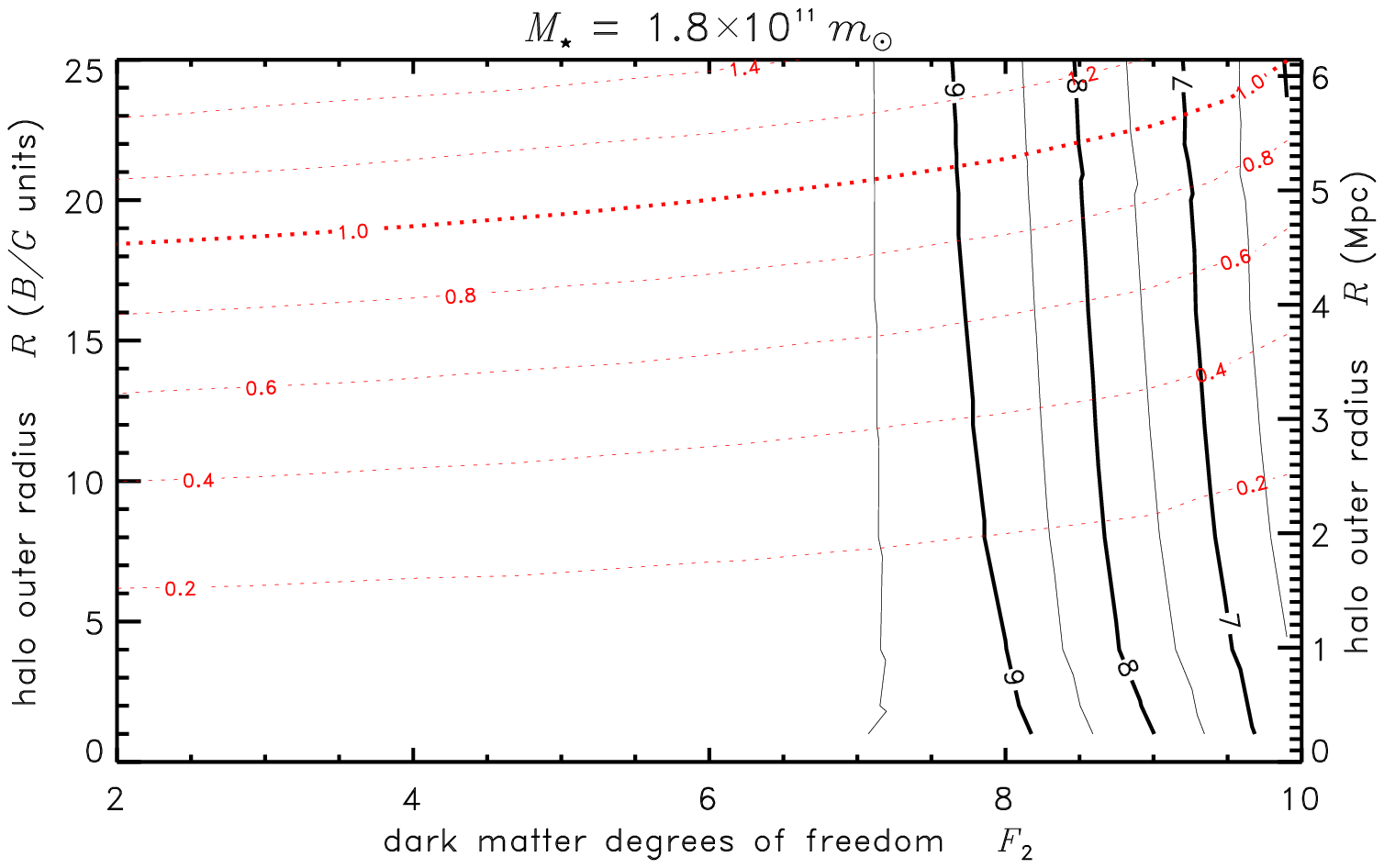}
\end{array}$
\caption{
Relations between cluster radius $R$,
   the dark matter degrees of freedom $F_2$
   and the minimum central mass (the mass for the central SMBH).
Black (solid) contours show logarithms of the minimal central mass,
   in solar units 
	($\log_{10}(m_*/m_\odot)$),
   found by minimizing over $(s_2,{\mathcal M}_R)$.
Red (dotted) curves are tracks where the baryon mass fraction is
   a certain multiple of a `cosmic' reference value
   ($0.16$).
The top panel shows the  clusters without a central cD galaxy
   (i.e.\ $M_\bigstar=0$).
The bottom panel shows the cluster containing
   a cD galaxy with $M_\bigstar=1.8\times10^{11}m_\odot$.
}
\label{figure.m.limit}
\end{figure}

Unsteady overcooling could cause a discontinuity to form at a certain radius. 
This can be avoided, 
  when a sufficiently massive central object is present as an accretor, 
  since heating due to the accretion process counteracts radiative cooling. 
For the parameters that we have considered,  
   all steady solutions have positive central mass, $m_* > 0$.
The main parameters to be explored are
   $(F_2,R,s_2,{\mathcal M}_R)$. 
By comprehensive numerical searches over $(s_2,{\mathcal M}_R)$, 
   we map the landscape of the minimal $m_*$ across the parameter space. 
Optimization routines can iterate to the best solution
   (within round-off)
   after integrating the model profiles
   at a few hundred ${\mathcal M}_R$ trial values,
   and a few hundred $s_2$ values for each of those.
Painstaking reiteration of this process at different $(F_2,R)$
   yields maps such as Fig.~\ref{figure.m.limit},
   where black contours show the minimal $m_*$
   and red shows cluster baryon fractions.

We previously
   explored the effects of $m(R)$, $T_R$, $\dot{m}$ and $F_2$
   upon the halo radius $R$ (for given baryon fraction)
   and for the possible values of
   the central point-like mass in the clusters without a cD galaxy
   \citep{saxton2008}.
Assuming a standard mass $m(R)$,
   these are the main findings.
\begin{enumerate}
\item
    When the baryon fraction is fixed, 
    larger $F_2$ gives smaller $m_*$, 
    i.e.\ a larger value for the internal degree of freedom
    of the dark matter particles 
    gives a small mass for the central point-like object.
\item
   Large gas inflow rate $\dot{m}$
   implies a larger limit for the central point-mass $m_*$,
   and greater sensitivity of $m_*$ to $F_2$.
\item
   For a given baryon mass fraction,
   the cluster halo radius $R$ decreases when $\dot m$ increases.
\item
   Raising the outer gas temperature $T_R$ reduces $R$,
   but leaves the $m_*$ map unchanged. 
\end{enumerate}
 
Observations have shown that  
   central black holes in many galaxies have masses  $\sim 10^6$--$10^9~m_\odot$
	\citep[see e.g.][]{gueltekin2009,graham2011}.
Our cluster models provide minimal $m_*$ in this range
   if the degrees of freedom of dark matter may be in the range $7 \la{F_2}<10$.
For smaller $F_2$, the predicted $m_*$ are excessively massive.
(Note that values of $F_2>10$
   cannot form a finite mass within a finite radius in our cluster model.)
This range of $F_2$ is consistent with polytropic halo fits
   to the scaling relations of disc galaxies,
   which imply $F_2\approx 9.6$
   \citep{nunez2006,zavala2006}.
Fitting the kinematics of planetary nebulae
   and stellar tracers in elliptical galaxies
   prefers a range $7\la{F_2}\la9$
   \citep{saxton2010}.
For our choices of $\sim1$~keV gas temperatures
   at the cluster outskirts,
   empirically plausible cluster halo radii of a few Mpc
   imply that gas inflows are more modest
   ($<100~m_\odot~{\rm yr}^{-1}$)
   than some X-ray observations implied.

Here, we consider a more general situation:
   the cluster contains a cD galaxy which provides
   a significant additional gravitational field
   in central regions,
   and the AGN within the cD galaxy provides radiative
   and/or mechanical heating to the intracluster gas.  
The central mass limits arise due to local thresholds of overcooling
   and supersonic catastrophes in the inner regions. 
This is affected by the central contraction of the halo by gas inflow
   (which depends on $F_2$).
However, since the stellar mass of a cD elliptical galaxy
   is dynamically significant at kiloparsec radii,
   it is desirable to test how it influences the domain of steady solutions.

We find that the inclusion of the cD galaxy
   (lower panel, Fig.\ref{figure.m.limit})
   leaves the cosmic composition tracks unchanged
   (red dotted contours).
This is somewhat unsurprising:
   in a more global view, the mass contribution of the cD galaxy
   is an insignificant detail compared to
   the overall mass distribution of the cluster.
The global properties of the cluster
   (e.g. mass, temperature and radius relations)
   are essentially unaffected by the localized perturbation
   in the innermost region of the cluster core.
    
However the presence of the cD galaxy has a subtle but notable effect
   on the central mass limit, $m_*$.
Without the galaxy (upper panel, black contours of Fig.~\ref{figure.m.limit})
   there is a large domain where the clusters with cosmic compositions
   always have a central object $m_*>10^{10}m_\odot$ if $F_2\la7$.
We infer $F_2\ga7$ on astronomical grounds.
When the galaxy is included, the $m_*$ limits
   drop by at most a factor $\sim 3$
   for cases where the halo has $F_2<7$
   (see the lower panel in Fig.~\ref{figure.m.limit}).
The mass limits for haloes with $F_2\ga7$ are not significantly affected
   (the tight black contours on the right-hand side of each panel).
Thus, the presence of a galaxy's stellar mass
   slightly lessens the observational exclusion of low $F_2$.
In the context of the SIDM model,
   the requirement of realistic central black holes
   still favours a larger number of thermal degrees of freedom
   ($F_2\ga7$).

We may interpret this finding as follows.
In effect,
   the stellar component is analogous to
   an extra halo component with shallow gradients.
It does not fundamentally change the local vulnerability
   of the gas flow to breakage in supersonic rips
   nor cooling catastrophes.
Nor does the galaxy affect the compressibility
   and dynamical characteristics of the DM.
The preference for high $F_2$ is a consequence of
   the dark matter's weak pressure response
   to density variations.
Steep central density gradients
   are required to provide pressure support in the interior,
   and those high densities enable a massive dark spike
   around a smaller point-mass.

\begin{figure}
\centering
$\begin{array}{cc}
\includegraphics[width=82mm]{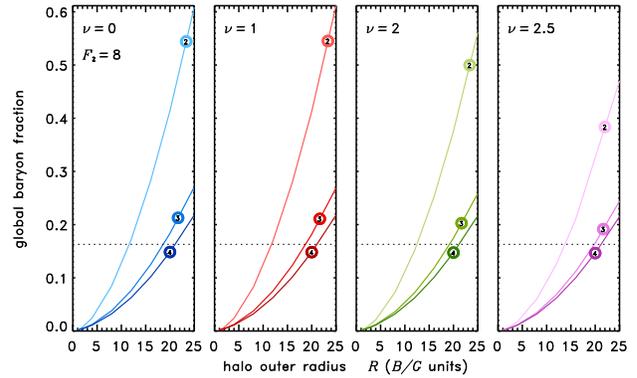}
\end{array}$
\caption{%
Baryon mass fraction as a function of halo radius ($R$)
   for clusters with a cD galaxy and spatially distributed heating by the AGN 
   ($Ar^{-\nu}$).
The cluster parameters are
   $F_2=8$, $T_R=1$keV, $\dot{m}=10m_\odot~{\rm yr}^{-1}$.
Curves indicate the heating power as labelled:
   $10^{-2}U_L, 10^{-3}U_L, 10^{-4}U_L$
   (where the fiducial luminosity
   $U_L\approx1.44\times10^{45}{\rm erg}~{\rm s}^{-1}$).
Panels from left to right show:
   spatially uniform heating ($\nu=0$);
   the a somewhat concentrated heater ($\nu=1$);
   the radiation-like heating ($\nu=2$);
   and a more concentrated model ($\nu=2.5$).
The dotted lines indicate a `cosmic' baryon fraction.
}
\label{figure.agn.frac}
\end{figure}

The inclusion of an AGN-like heating function ($A>0$)
   has a moderate effect on the size relations among clusters.
Figure~\ref{figure.agn.frac}
   shows the results of different heating rates
   upon minimal $m_*$ clusters of given radius ($R$),
   for fixed inflow rate ($\dot{m}=10\,m_\odot~{\rm yr}^{-1}$),
   gas surface temperature ($T_R=1$keV),
   total mass
   and halo physics ($F_2=8$).
If the heating is of the order of the fiducial power
   ($U_L\approx1.44\times10^{45}~{\rm erg}~{\rm s}^{-1}$),
   then we find no stationary solutions for clusters
   with this choice of $(\dot{m},m,T_R,F_2)$.
At a given baryon fraction,
   heating at rates of $10^{-2}U_L$
   results in a smaller cluster radius ($R$)
   than in unheated clusters
   (upper curves of Fig.~\ref{figure.agn.frac}).
If the heating is reduced to $\la10^{-4}U_L$,
   then the size and composition relations
   are indistinguishable from unheated models
   (lower curves of Fig.~\ref{figure.agn.frac}).
Comparing the panels of Fig.~\ref{figure.agn.frac} shows that
   the global rate of heating is the most decisive parameter;
   the radial index of the heating function makes little difference
   in the domain we explored
   ($\nu=0$, 1, 2 and 2.5).
For larger $\nu$ (more concentrated heating),
   the AGN is less effective at reducing cooling
   at cluster core scales,
   and the results become marginally less sensitive
   to the power parameter $A$.
Qualitatively, non-gravitational heating
   enables denser gas profiles to avoid local overcooling catastrophes;
   the `too cold' zone occupies less of the configuration-space.
At a fixed cluster size $R$, greater gas fractions are possible.
At a fixed gas fraction, clusters can be more compact.
Heating is more effective if it occurs in the kpc-scale cool core,
   less effective if it concentrates in the hot inner nucleus.

\begin{figure}
\centering
$\begin{array}{cc}
\includegraphics[width=82mm]{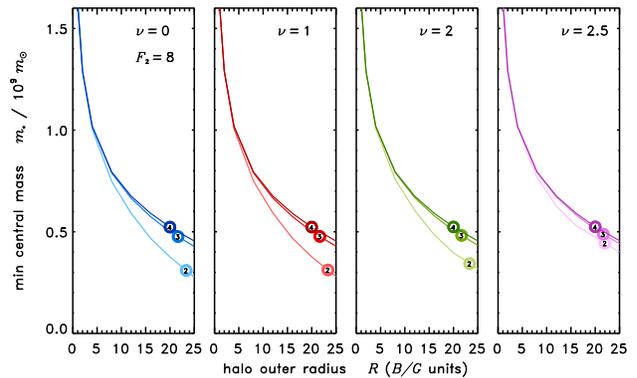}
\end{array}$
\caption{%
Minimal central point-mass ($m_{*}$) as a function of halo radius ($R$)
   for clusters with a cD galaxy and spatially distributed heating.
The cluster parameters and heating power of the AGN 
   correspond to the panels and curves in Fig.~\ref{figure.agn.frac}.
}
\label{figure.agn.mass}
\end{figure}

Whether or not heating affects the cluster size relations
   is independent of the halo microphysics.
We confirmed this by calculating $F_2=3$ results
   equivalent to the $F_2=8$ models
   illustrated in Fig.~\ref{figure.agn.frac}.
However, the heat sensitivity of the minimal central mass ($m_*$)
   does depend on the dark matter degrees of freedom.
Fig.~\ref{figure.agn.mass}
   shows how strong heating can lower the limiting $m_*$ 
   by up to a few tens of percent when $F_2=8$.
When there are fewer degrees of freedom
   (e.g. $F_2=3$ calculations)
   the $m_*$ versus $R$ curves differ negligibly
   between $10^{-2}U_L, 10^{-3}U_L, 10^{-4}U_L$ and unheated clusters.
The same is found when the polytropic halo
   is replaced by the popular but cuspy `NFW' profile
   of collisionless DM simulations
	\citep{nfw1996}.
Appendix~\ref{s.nfw} illustrates and summarizes
   the results of these variants.
The central mass and global scaling properties of NFW clusters
   resemble those of $F_2\approx3$ clusters:
   the minimal $m_*$ is ultramassive,
   and AGN heating has negligible effect compared to
   the introduction of the cD galaxy potential.

\subsection{Density profiles}

\begin{figure}
\centering
$\begin{array}{cccc}
\includegraphics[width=82mm]{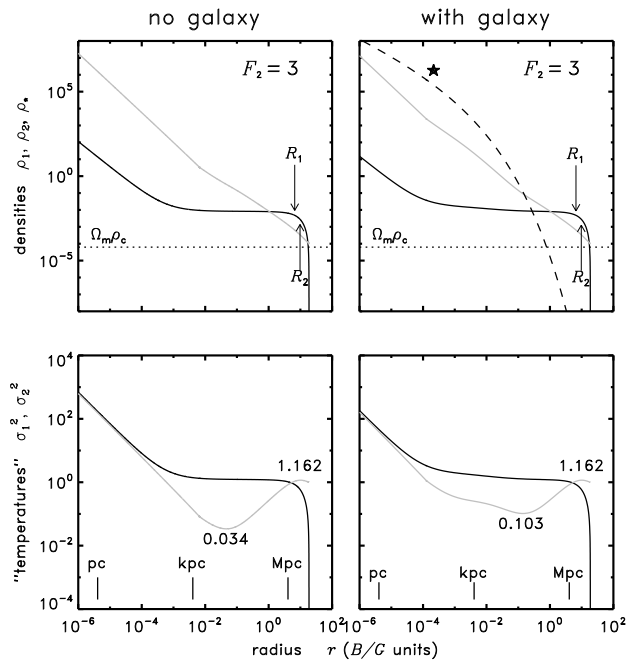}
\\
\end{array}$
\caption{
Radial profiles of density (top row),
   and `temperatures' or velocity dispersions (bottom)
   for minimal $m_*$ clusters.
Gas is shown grey; dark matter in solid black curves.
The dotted line is the standard cosmic mean density;
   the dashed curve is the stellar density.
These particular solutions
   were chosen as those with the minimum central mass,
   and thus a large ICM temperature variation.
This figure shows cases with $F_2=3$
   dark matter degrees of freedom,
   analogous to a particles of a monatomic plasma.
In the left-hand column there is no cD galaxy
   \citep[as in][]{saxton2008};
   in the right-hand column the fiducial galaxy has been added.
Arrows mark $R_1$ and $R_2$,
   indicators of core sizes in terms of the density slope
   (${\mathrm d}\ln\rho/{\mathrm d}\ln r=-1,-2$ at $R_1, R_2$).
The gravitational effect of the cD galaxy raises the gas floor temperature
   by a factor $\approx3$.
Local extrema of the gas temperature are annotated.
The three ticks in the bottom panels
    show the $1$pc, $1$kpc and $1$Mpc radii.
}
\label{figure.radial}
\label{figure.radial.F3}
\end{figure}

\begin{figure}
\centering
$\begin{array}{cccc}
\includegraphics[width=82mm]{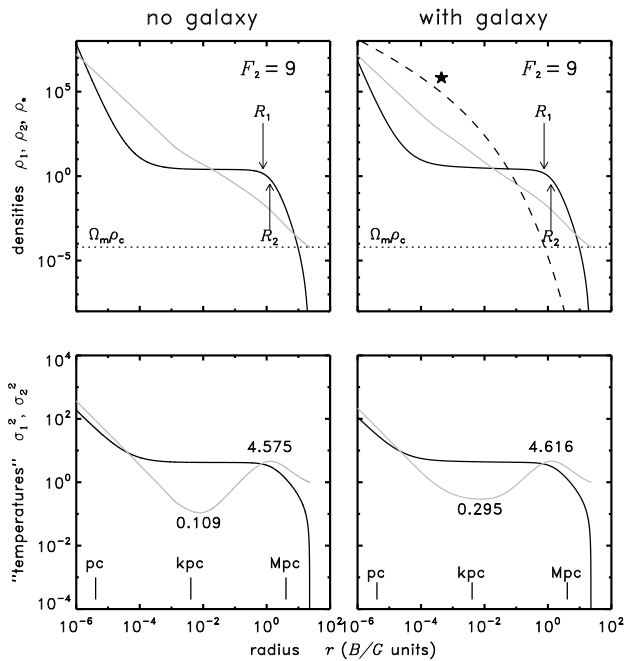}
\\
\end{array}$
\caption{
Density and velocity dispersion profiles
   for minimal $m_*$ cluster models
   as in Fig.~\ref{figure.radial.F3}
   but now with greater dark matter heat capacity,
   $F_2=9$.
The dark matter core is smaller.
Like the $F_2=3$ case,
   the floor temperature of the gas is raised by a factor of $\approx3$.
}
\label{figure.radial.F9}
\end{figure}

We generally obtain a power-law-like gas density profile
   with the power-law index $\approx1$
   (top row, Figs~\ref{figure.radial.F3} and \ref{figure.radial.F9}). 
This is similar to the profiles in the conventional `cooling flow' models
	\citep[see e.g.][]{cowie1977,fabian1977}.    
There is usually a subtle break in the slope at a kpc radius or smaller.
The innermost gas profile below a few pc
   is $\rho_1\sim r^{-1.5}$,
   like a \cite{bondi1952} accretion spike.

The DM has a more complicated density profile. 
A parsec-scale dark density spike
   surrounds the central mass.
This is not accreting material like the gas cusp,
   but an adiabatic structure supported by its own pressure.
This inner subsystem is a gravitational domain-of-influence
   belonging to the central mass
   and the spike material itself
\citep[similar to e.g.][]{huntley1975,quinlan1995,
	gondolo1999,ullio2001,macmillan2002}.
Even if dark matter were collisionless in extragalactic conditions,
   it becomes effectively collisional in the nuclear region,
   via scattering with stars there
	\citep{ilyin2004,gnedin2004a,merritt2004,zelnikov2005b,
		vasiliev2008,merritt2010}.
In the Newtonian (weak gravity) regions of the spike,
   the dark density profile depends on
   the number of effective thermal degrees of freedom,
   $\rho_2\sim{r}^{-F_2/2}$.
Now we find that adding the stellar mass profile of a cD galaxy
   has no significant effect on the occurrence of spikes,
   nor the spike's density gradient.
However, for the minimal $m_*$ solutions,
   the cD galaxy environment makes the spike an order of magnitude less dense
   (and lower velocity dispersion by a smaller factor).

Outside the spike,
   the density is almost uniform 
   throughout a core spanning 10~kpc to several 100~kpc
   for plausible cluster models.
Outside this core,
   the dark density fringe declines smoothly
   towards zero at the truncation radius $R$.
We might define the edge of the dark core
   as the radius where the density index passes specific values
   (e.g. ${\mathrm d}\ln\rho/{\mathrm d}\ln{r}=-2$ at $r=R_2$,
   or slope $-1$ at $r=R_1$).
These locations are marked with arrows in
   Figs~\ref{figure.radial.F3} and \ref{figure.radial.F9}.
Similarly to single-fluid polytropes,
   the size of the cluster's dark matter core
   depends on $R$ and the dark matter degrees of freedom, $F_2$.
In the minimal $m_*$ solutions shown in
   Figs~\ref{figure.radial.F3} and \ref{figure.radial.F9},
   the core is clearly smaller when $F_2=9$ than when $F_2=3$.
For greater degrees of freedom, the core is generally smaller.
The inclusion of a kpc-sized central galaxy
   only shrinks the dark core by $\la1$\%.
For our standard clusters with cD galaxy and cosmic composition,
    when $F_2=8, 9, 9.5$ and $9.9$ the core sizes give
    $R_1\approx366, 187, 98.5, 21.7$~kpc, respectively.
In hindsight, our standard models seem
   a bit diffuse and colder than observed clusters.
This is easily remedied:
   the radii would shrink if $|\dot{m}|$ or $T_R$
   were increased at the outer boundary.
Alternatively, under the scaling homologies,
   the model is equivalent to a cluster
   a few times hotter and more massive than standard $4\times10^{14}~m_\odot$
   (whilst keeping the same radial dimensions).

\subsection{Temperature profiles}

If self-gravity and the kinetic terms in the hydrodynamics are ignored
   (in the conventional `cooling flow' models),   
   gas temperatures in the inner core region of a cluster can reach zero
   when runaway radiative cooling develops.  
As a result, the ratio of peak to coolest gas temperatures
   $T_{\rm max}/T_{\rm min}$ is expected to be very large throughout the core,
   which is inconsistent with the observations that
   $T_{\rm max}/T_{\rm min} \sim (2$--$5)$ for most cool-core clusters
	\citep{kaastra2001,tamura2001,peterson2001,boehringer2001,
		sakelliou2002,ettori2002,
		peterson2003,voigt2004,bauer2005,zhang2006,werner2006,
		fujita2008,takahashi2009,bulbul2012}.
We demonstrate that when self-gravity and kinetic terms
   are included in the hydrodynamics,  
   the gas is a cooling-induced inflow
   and the ICM floor temperature ($T_{\rm min}$) is non-zero.
This is due to the fact that the radiative cooling of the intracluster gas
   is counterbalanced by accretion warming, 
   averting the runaway cooing process. 
The accretion effect tends to be larger for larger $m_*$
    or when a cD galaxy is present.
Although runaway cooling can occur (in the `too cold' configurations),
   in more realistic settings $T_\mathrm{min}$ is non-zero,  
   and the finite value of the ($T_\mathrm{max}/T_\mathrm{min}$) ratio  
   depends on the gas inflow parameters
   and other global parameters of the clusters \citep{saxton2008}.  

The lower panels of
	Figs~\ref{figure.radial.F3} and \ref{figure.radial.F9}
   depict temperature profiles
   of some minimal $m_*$ clusters with and without the cD galaxy,
   for $F_2=3$ and $F_2=9$ haloes,
   in cases matching the standard cosmic baryon fraction.
Gas temperature profiles are generally S-shaped curves:
   hottest around the central accretor;
   a minimum ($T_\mathrm{min}$ at $R_\mathrm{min}$)
   farther out in the `cool core';
   a peak ($T_\mathrm{max}$) at $R_\mathrm{max}\sim$Mpc radii;
   then a decline in the outer fringe.
The coldest layer of the cool core
   occurs at radii $R_\mathrm{min}$ of a few kpc to tens of kpc.
The minimal $m_*$ model always provides
   the most extreme ICM temperature decrement
   (greatest $T_{\rm max}/T_{\rm min}$),
   other solutions show less temperature variation.
All else being equal, $R_\mathrm{max}$ is smaller if $F_2$ is larger.
Otherwise this peak location is mainly dependent on the cluster radius $R$,
   but somewhat affected by outer boundary properties
   ($\dot{m},\Mcal_R,T_R)$.
The floor $T_\mathrm{min}$ is controlled by
   accretion and heating in central regions.
Adjusting these quantities
   can provide fits to observed X-ray temperature profiles,
   comparable to the NFW-based parametric fits of \cite{vikhlinin2006}
   (but without needing to omit data in the inner tens of kpc).
See Appendix~\ref{s.fitting} for examples.
Each physically self-consistent fit
   entails exploration of a large parameter-space,
   which is best left for a dedicated paper.

Fig.~\ref{fig.TmaxTmin} shows the relation between
   $T_{\rm max}/T_{\rm min}$ and $m_*$
   for various types of clusters
   with different AGN and cD conditions
   (but all with $F_2=8$ and $R=16U_x\approx4$~Mpc identically).
When there is no cD galaxy in these clusters
   the ($T_{\rm max}/T_{\rm min}$) ratios are $\sim 1 - 40$
   (black dashed curve in Fig.~\ref{fig.TmaxTmin}).  
The relation between ($T_{\rm max}/T_{\rm min}$) and $m_*$
   is not easily described in terms of a simple function.
It is non-monotonic if $F_2>6$.
However, we can see that the ($T_{\rm max}/T_{\rm min}$)
   ratio can be reduced when $m_{*}$ is sufficiently large. 
Thus, in principle $(T_{\rm max}/T_{\rm min}) \approx 1$
   can be attained for very large $m_{*}$.
In such situations, 
   there is no temperature minimum;
   the ICM becomes continually hotter nearer the centre
   (like a non-cool-core cluster).

When a cD galaxy is present in the cluster,
   its stellar mass strengthens the power generation via accretion, 
   counteracting the radiative cooling of the gas in the cluster core region.  
This raises $T_{\rm min}$ substantially at $r\la30$~kpc
   and hence reduces the ($T_{\rm max}/T_{\rm min}$) ratio   
   (grey curve in Fig.~\ref{fig.TmaxTmin}).   
The stellar mass profile of the cD galaxy
   distributes the accretion power smoothly over regions at kpc radii, 
   rather than concentrated in compact region around the central point-mass.
The temperature floor is raised accordingly,
   creating a softer temperature gradient 
   in the intracluster gas around $\sim10^1$~kpc radii. 
Note that the presence of the cD galaxy may boost $T_{\rm min}$ by some factors  
  (about three for minimal $m_*$ cases),
  but $T_{\rm max}$ does not change significantly.

Next, we consider the addition of heating by an AGN in the cD galaxy.  
We assume that the AGN heating is steady, 
   as the duty cycles of AGN may be on time-scales of million years 
   and the global dynamical time-scales of the cluster
   are about hundred million years.
We parametrized the distribution of AGN power in the intracluster gas 
  by means of spatial power laws with various indices, $\nu=0, 1, 2$ and 2.5.
Our calculations have shown that
   for $F_2=8$ clusters in the presence of a cD galaxy, 
   the attainable ranges of $T_\mathrm{max}/T_\mathrm{min}$
   are very similar for different types of AGN distributed heating 
   (the colour curves in Fig.~\ref{fig.TmaxTmin}). 
Moreover, they are also very similar to the case with an inactive AGN
   (i.e.\ just a cD galaxy). 
For $F_2=8$ clusters,
   the curves of different $\nu$ are similar but displaced in $m_*$.
For $F_2=3$ or clusters with the `NFW' halo profile,
   the AGN heating increases the $T_\mathrm{max}/T_\mathrm{min}$ ratio
   at any given $m_*$
   (Appendix~\ref{s.nfw}).
Interestingly,
   the AGN-heated $T_\mathrm{max}/T_\mathrm{min}$-$m_{*}$ curves 
   approach the curve for an inactive galaxy
   when the value of $\nu$ increases.  
Nevertheless, the heating distribution index $\nu$ influences
   the minimum value of $m_{*}$ possible in the model.
This is not too surprising as  
   centrally concentrated AGN heating and accretional heating
   via the presence of cD galaxy 
   share a similarity:
   the heat generation in both cases
   is in the core region of the cluster. 

In summary,  the temperature floor $T_{\rm min}$ can be raised 
   essentially by the presence of some heating sources
   to counterbalance the radiative losses of the gas.  
The heating sources can be hydrodynamical heating
   (via accretion on to a point-mass or cD galaxy)
   and/or power from an AGN.    
Accretion heating via the presence of a cD galaxy is able to produce 
    $T_{\rm max}/T_{\rm min}$ 
    compatible with presently available observations.  
The distributed heating of an AGN may lower the $T_{\rm max}/T_{\rm min}$ ratio 
    but appears less influential than the cD galaxy's gravity.
Nevertheless, the lower limit for $m_{*}$ depends on 
   how the AGN power is distributed spatially
   and/or on whether or not an AGN is present.

\section{Discussion}
\label{s.discussion}

\subsection{Symmetry and extra gas processes}

This paper emphasizes the role of the cD galaxy in a cooling ICM
   in a naturally cored galaxy cluster.
Our treatment of the gas does not contain all conceivable processes,
   and we should briefly consider the relevance of extra physics.

We warn that our findings might not be relevant for systems
   that violate the spherical, isotropic, stationary assumptions.
The constraints might change in a real cluster
   that is distorted by non-spherical or anisotropic effects:
   bulk rotation,
   streaming motions,
   massive substructures,
   gross departures from spherical symmetry,
   local anisotropies and non-thermal plasma effects,
   or magnetically dominated pressure.
Our models are also inapplicable to significantly non-stationary clusters,
   e.g. during violent mergers.
Such convulsions would of course induce global mixing and shocks,
   temporarily erasing the cooling gas inflows
   that we intended to explain.

Among the smaller scale physics,
   we intentionally omit the mass sink terms
   (describing condensation of dense cold blobs from the ICM)
   that feature in other cooling flow models
   \citep[especially at galaxy scales, e.g.][]{mathews2003}.
Our justification is empirical:
   the cool condensates and star formation
   are not observed in the required amounts
   \citep{donahue2004}.
The inflowing matter must have a different fate,
   deep in the nuclear regions.
We also omit
   potential complications, such as
   radiation pressure,
   magnetic field geometry
   and rotation,
   which could determine the detailed internal anatomy of the AGN
   at sub parsec radii.
Our goal is to explain the cluster structure
   without depending too much on microscopic complications.

Gas turbulence could boost the effective internal heat capacity of the ICM,
   due to the kinetic energy density of eddies.
Observations suggest that this does not add more than
   a few tenths of the thermal pressure component
   \citep[e.g.][]{sanders2009}
   which would imply $3<F_1<4$ (and closer to the lower limit).
A slightly enlarged $F_1$
   may defer overcooling in particular models,
   loosening the constraint
   of the `too cold' zone in configuration space
   (but probably not enough to make a qualitative difference
   to $m_*$ limits and cluster compactness).

The fluid description depends on the mean free path (mfp) of ICM particles
   ($\propto \sigma^4/\rho$)
   being much smaller than $r$.
Processes that depend on scattering
   (especially the free--free cooling function)
   are effectively blurred across that length-scale.
Overcooling catastrophes might be averted in some marginal cases
   (relaxing constraints in the configuration space).
Such a semicollisional model is not expressible
   as an initial value problem of ODEs,
   and would need a different numerical scheme.
Fortunately,
   for our most realistic cluster solutions
   (in terms of $m_*$ and size)
   the estimated mfp in the cool core and Mpc outskirts
   turns out to be at most a few percent of $r$,
   justifying the fluid approximation.
Only in the nuclear region ($r\la0.1$~pc)
   does the gas become collisionless in mfp terms.
Since the overcooling catastrophes occur at kpc radii,
   this aspect of AGN anatomy will not affect the key conclusions of this paper.

Our present formulation omits thermal conduction.
This process would introduce another energy exchange term in $\Lcal_1$,
	\citep[$\propto\nabla\cdot(\sigma_1^5\nabla\sigma_1^2)$;
	][section~5.4.2]{sarazin.book}
   and require specification of an outer boundary condition on
   $\nabla\sigma_1^2$.
Qualitatively,
   we would expect conduction to soften temperature gradients,
   by spreading warmth from the inner accretion region
   outwards to the cooling core.
Previous works emphasized conductive heating of the core by the outskirts
   \citep[e.g.][]{narayan2001,ruszkowski2002,voigt2004,ruszkowski2010}.
The ICM temperature floor $T_\mathrm{min}$
   might be raised further than in the present models,
   and the highest $T_\mathrm{max}$ values would also decrease.
Realistic temperature ranges
   ($T_\mathrm{max}/T_\mathrm{min}<5$)
   might occur for a broader domain of cluster parameters.
In the inner sub parsec accretion region,
   conduction could dominate over emission,
   dimming the central X-ray source.
Such effects are probably insensitive to $F_2$ halo physics
   and cD galaxy profile,
   but may make an interesting topic for future modelling.

\begin{figure}
\centering
$\begin{array}{cccc}
\includegraphics[width=84mm]{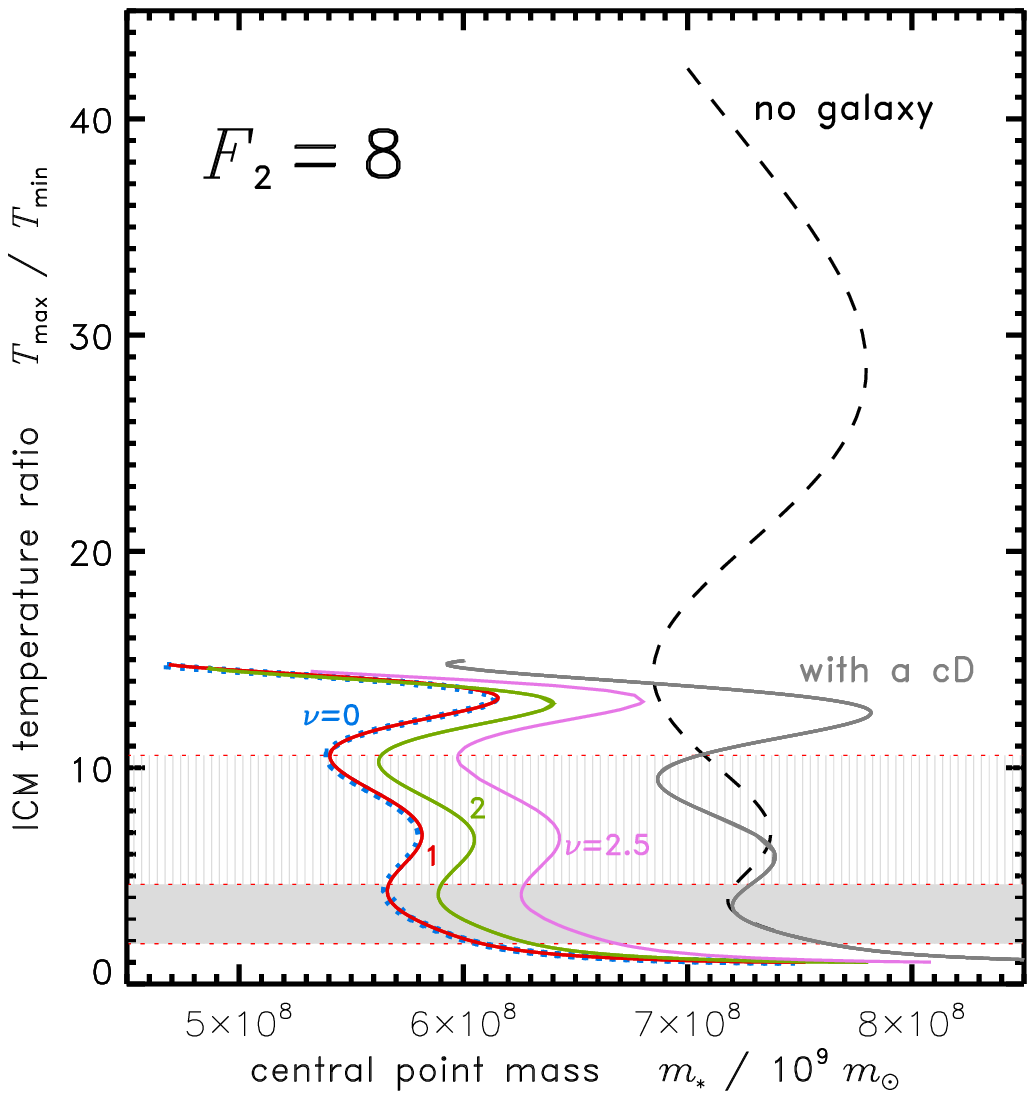}
\\
\end{array}$
\caption{
The ($T_\mathrm{max}/T_\mathrm{min}$) ratio of the intracluster gas 
   for cluster models with $F_2=8$ and radius $R=16$,
   along the borders of the {\it `deep zone'} continuous models.
The horizontal axis is the central point-mass $m_{*}$.
The dashed black curve shows the ratios for clusters without a cD galaxy.
The grey curve shows results with an inactive cD galaxy. 
Coloured curves show the ratio for the clusters with an AGN in the cD galaxy. 
The power of the AGN is $10^{-2}U_L$
   and the spatial indices of the AGN heating functions are
   $\nu=0,$ 1, 2 and 2.5 respectively. 
The grey shaded band shows the normal range of cool-cored clusters;
   the lined region includes Centaurus.
The gravity of the cD galaxy
   is able to significantly lower the ($T_\mathrm{max}/T_\mathrm{min}$) ratio.
The form of heating function
   affects the attainable value of central point-mass $m_{*}$.
}
\label{fig.TmaxTmin}
\end{figure}

\subsection{Jeans stability and dark collapse}

\begin{figure}
\centering
$\begin{array}{cc}
\includegraphics[width=82mm]{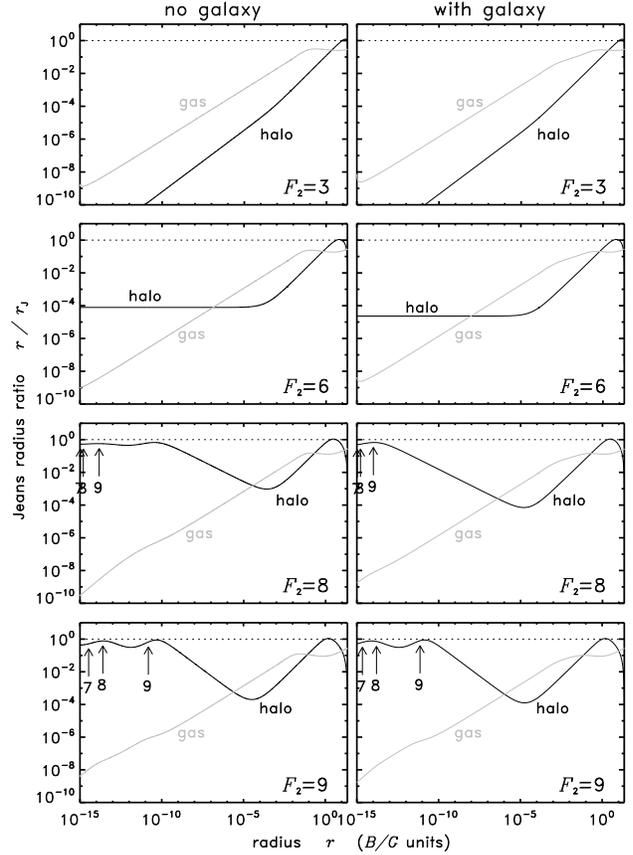}
\end{array}$
\caption{
Radial profiles of Jeans stability ratio $r_{\rm J}/r$
   for some minimal $m_*$ cluster solutions
   in haloes with various $F_2$ values (annotated)
   and total baryon fraction $\approx0.16$.
The dotted line shows the margin for instability.
Grey curves depict the gas profile;
   black curves depict the dark matter.
In the left-hand column, the model contains gas and dark matter only.
In the right-hand column, a cD galaxy is also included
   ($M_\bigstar=1.8\times10^{11}m_\odot$).
Numbered arrows mark the radii that contain
   $10^7, 10^8$ and $10^9m_\odot$ of dark matter.
}
\label{figure.jeans}
\end{figure}

As an indicator of local gravitational stability,
we can calculate a Jeans radius at any layer of the cluster:
\begin{equation}
	r_{\rm J} \equiv \sqrt{{\uppi\gamma\sigma^2}\over{4G\rho}}
	\ .
\end{equation}
Stability would be indicated when the Jeans radius is greater than
   the radius of a particular layer locally, $r_{\rm J}>r$.
In the absence of a cD galaxy,
\cite{saxton2008}
   found the gas to be stable everywhere,
   and increasingly stable in the interior
   (grey lines in the left-hand panels of Fig.~\ref{figure.jeans}).
Now we find that the addition of a cD galaxy stellar mass profile
   can subtly flatten the profile of $r_{\rm J}/r$ for gas,
   at radii comparable to $R_{\rm e}$
   (see the grey curves of the right-hand panels of Fig.~\ref{figure.jeans},
   and compare to the upper panels).
By this criterion, the gas inflow remains stable
   in the presence of the galaxy, just as it was without a cD galaxy.

The corresponding property of the dark halo
   depends on the thermal degrees of freedom, $F_2$.
For cases with $F_2<6$
   the central regions of the halo are increasingly stable,
   similarly to the gas.
However for $F_2>6$ the dark spike
   rises towards the threshold of instability
   ($r_{\rm J}\approx r$) in a deep interior within $r<10$~pc.
This is a manifestation of the classic
   dynamical instability of pure polytropic spheres with high index
   \citep[][p.~51]{ritter1878,emden1907,chandrasekhar1939}.
Simplistically, for large $F$ the pressure response
    to a local perturbative compression
    may be insufficient to restore dynamical equilibrium.
We find that the $r_\mathrm{J}/r$ behaviour persists
   after introducing the distributed stellar mass of a cD galaxy.
Comparing minimal $m_*$ models with and without the galaxy,
   the radial gradients of $r_{\rm J}$ are only subtly changed.
This phenomenon occurs at such small radii
   that the introduction of the stellar background mass distribution
   --- mainly farther out at kiloparsec scales
   --- makes no significant effect.

The upturn of $r/r_{\rm J}$ for the inner part of the dark halo
   implies that external perturbations could cause it
   to detach and collapse on its dynamical time-scale.
Gas does not participate directly,
   so the implosion is dark, evading the limits of
   \cite{eddington1918b} and \cite{soltan1982}.
Arrows in the lower panels of Fig.~\ref{figure.jeans}
   indicate radii enclosing $10^7, 10^8$ and $10^9m_\odot$.
For cluster-sized solutions and large $F_2$,
   the fragile part of the halo is up to a billion solar masses.
This implies that an SMBH
   could grow substantially in a `dark gulp'
   \citep{saxton2008}.
This process may alleviate the need for
   conspicuously luminous gas accretion,
   and explain the rapid formation of SMBH
   implied by the existence
   of powerful quasars at high redshifts, $z>6.4$
   \citep[e.g.][]{fan2004,mortlock2011}.
Dark gulping may be a factor counteracting
   a more modern issue of black hole demographics:
   it is thought that some merged black holes
   can eject from the host galaxy due to gravity-wave recoil
   \cite[e.g.][]{haiman2004,baker2006,campanelli2007a,campanelli2007b,
	schnittman2007,lousto2011,lousto2013}.
Our scenario suggests that a replacement SMBH
   could condense naturally at the centre of the dark halo
   \citep[e.g. as in NGC~1275;][]{shields2013}.
It is reassuring to confirm that
   this `gulping' prediction
   is essentially unchanged by the cD galaxy's stellar profile.
If the such events can be prevented
   (in the high-$F_2$ SIDM context)
   it would require additional non-gravitational physics.
The persistence of collisionless stars in their orbits
   can have a stabilizing effect
   on the surrounding polytropic halo
   \citep{saxton2013}.
Details of the collapse modes must depend non-trivially upon
   the stellar and dark density profiles in the nuclear region
   \citep[as in the `skotoseismology' of gasless galaxies;][]{saxton2013}.

\section{Conclusions}
\label{s.conclusions}

We generalize our previous work on hydrodynamical structures
   and stability properties of stationary,
	locally isotropic,
   spherical galaxy clusters.
This basic model included self-gravity and the kinetic terms, 
   which were missing from many conventional analyses
   of hydrodynamics of galaxy clusters, 
   enabling us to correct and improve upon
   past understandings of cooling-induced gas inflows.  
Combined with observational constraints,
   our study gives insights into the dark matter physics.
In this work, we investigate the effects
   of the presence of a cD galaxy and central AGN.
The stellar mass profile of a cD galaxy exerts a gravitational influence
   in the cluster interior, 
   and the AGN provides distributed heating into intracluster gas.

We obtain exhaustive constraints on the profile
   and central object of a galaxy cluster with gas inflows,
   within the scope of our assumptions and with the stated ingredients.
Our results depend on the stationarity, spherical symmetry
   and isotropy of the cluster and the active central galaxy.
The constraints might be eluded,
   to an unknowable extent,
   if these conditions are unmet.
With that caveat,
   we summarize our findings as follows.
\begin{enumerate}
\item
The cD galaxy provides additional accretion warming 
   (due to stellar mass distributed at kiloparsec scales). 
   This raises the floor temperature of the intracluster gas 
   and hence reduces the temperature gradients in the cool core of the cluster. 
   Hence, a finite temperature floor in the intracluster gas can be attained  
      in the presence of a cD galaxy 
      regardless of distributed heating by an AGN
      or other non-gravitational heat sources.
\item
For given intracluster gas and dark halo masses,
    gas inflow rate and external temperature,
    minimal $m_*$ clusters with AGN distributed heating exceeding
    $\ga10^{41}~{\rm erg}~{\rm s}^{-1}$
    are slightly more compact than equivalent unheated models.
\item
Smaller dark matter cores (less than a few hundred kpc)
   still require large degrees of freedom $F_2$ for the dark matter.
     On these grounds alone, the preferred range is $6\la{F_2}<10$.
   The presence of a cD galaxy does not significantly modify
   the structures of the outer halo fringe,
   nor the inner core of the DM.
    Also it does not affect the mass spike surrounding the central object.
\item
Stationarity of the cluster demands a positive minimal central point-mass
     regardless of the cD galaxy or its AGN. 
   The range of the observed masses of SMBH in galaxies
      suggests that $7\la{F_2}<10$.
   The presence of a cD galaxy loosens the lower limit on realistic $F_2$ values,
       but not greatly.
    AGN distributed heating lowers the $m_*$ limit, 
       when the dark matter particles have large degrees of freedom
    (e.g. $F_2=8$).  
    The dark haloes of these clusters have large heat capacities. 
    The AGN effect upon $m_*$
    is insignificant when $F_2$ is small (e.g.\ $F_{2}=3$).
\item
The value of $m_{*}$ can be affected by how the AGN power
    is spatially distributed in the intracluster gas. 
     If $F_2$ is large, the minimum values for the $m_*$ limit
     are smaller for more evenly distributed AGN heating.
    If $F_2$ is small or the halo has NFW form,
    the radial index of the heating does not affect $m_*$ significantly.
\item
The temperature variation within the ICM ($T_\mathrm{max}/T_\mathrm{min}$)
     is maximal when $m_*$ is minimal.
     Introducing the cD galaxy shrinks this ratio by a factor $\sim3$.
     A sufficiently large point-mass $m_*$ reduces
     $T_\mathrm{max}/T_\mathrm{min}$ to the observed range ($\la5$).
    For large $F_2$ haloes these $m_*$ values are consistent with real SMBH;
    for $F_2=3$ or NFW haloes an ultramassive object is required.
\item
The presence of a cD galaxy does not significantly affect
     the stability of the inner gas profile,
     and the inner dark halo remains prone to gravitational collapse
     in spite of the presence of a cD galaxy.   
     Thus, a central SMBH can still condense
     or feed in dark gulps.
\end{enumerate}

\section*{acknowledgements}
This research has made use of NASA's Astrophysics Data System.

\bibliographystyle{mn2e}

\appendix

\section{alternative halo profiles}
\label{s.nfw}

Our main results focused on cluster models with a large $F_2$ halo,
   because these give $m_*$ mass limits
   consistent with realistic SMBH.
Here, we will briefly compare the benchmark $F_2=8$ models
   to the results
   for point-like `monatomic' SIDM ($F_2=3$)
   and the popular `NFW' profile
   \citep{nfw1996}.
`Monatomic' SIDM ($F_2=3$)
   can represent conditions similar to the most often studied SIDM theories
   \citep[e.g.][]{moore2000,ahn2005,peter2012,rocha2012}
   but with stronger scattering.
The NFW profile could describe the cuspy or small-cored haloes
   of collisionless DM theories,
   or forms of SIDM that are so weakly collisional
   that the core fails to grow to the full size
   attainable in an ideal Lane--Emden sphere.
The NFW radial density profile is
\begin{equation}
	\rho_\mathrm{nfw}={{\rho_0 R_\mathrm{s}^3
		}\over{
	r (R_\mathrm{s}+r)^2}}
	\ .
\label{eq.nfw}
\end{equation}

For the NFW modelling,
   we select a representative concentration
   $c=R_\mathrm{v}/R_\mathrm{s}\approx3.471$
   from the mass--concentration relation of \cite{duffy2008}.
The corresponding `virial radius' and scale radius are
   $R_\mathrm{v}=5.897U_x\approx1.45$Mpc
   and $R_\mathrm{s}=1.699U_x\approx0.42$~Mpc.
The halo is truncated at the chosen radius $R$.
For each trial choice of the outer Mach number
   (${\mathcal M}_R$) and temperature ($T_R$),
   the NFW density normalization ($\rho_0$) is varied until
   the total mass of the system
   (gas + stars + halo + SMBH)
   matches our standard value
  ($40U_m$).
This density scale replaces the role of $s_2$
   as a computational search parameter
   in the cluster configuration-space.

In all cases,
   the gas inflow from external cosmic background
   is set to our standard $T_R=1$keV temperature
   and $\dot{m}=10~m_\odot~\mathrm{yr}^{-1}$.
Under the effect a cD galaxy and AGN heating,
   the global mass-radius relations
   and the baryon fraction versus $R$ relations
   of the $F_2=3$ and NFW models are qualitatively similar to those of $F_2=8$,
   and need not be shown here.

\begin{figure}
\centering
$\begin{array}{cccc}
\includegraphics[width=84mm]{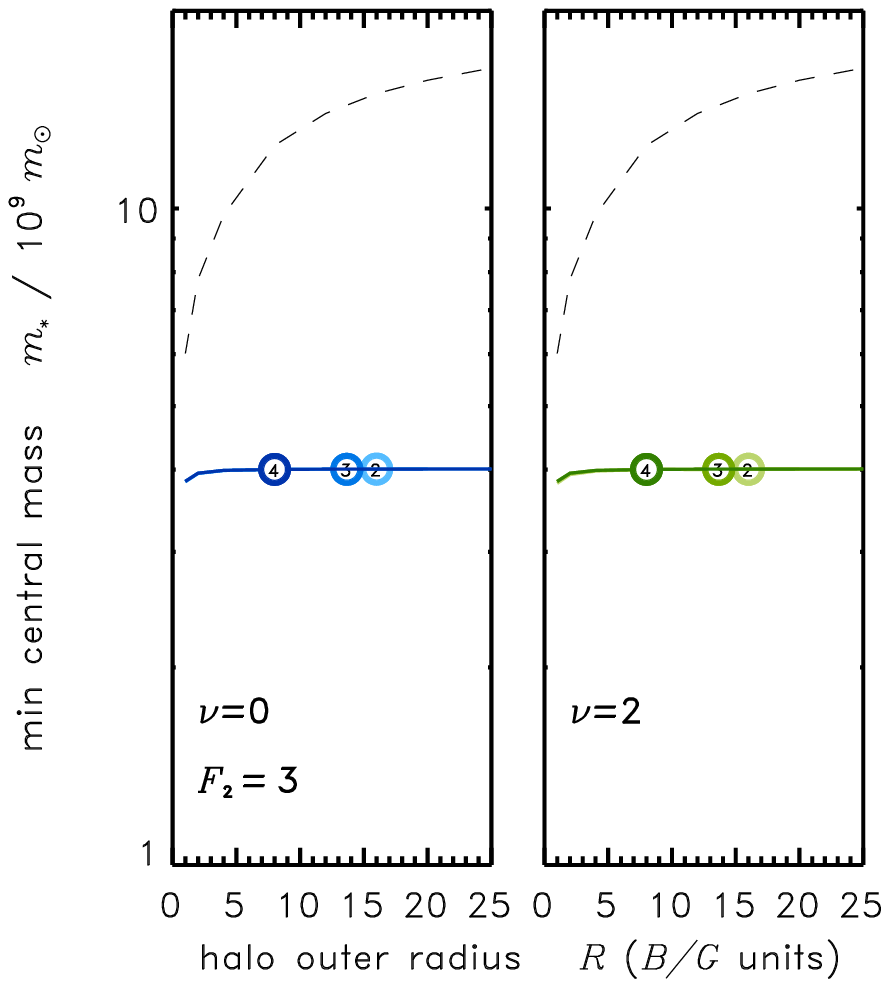}
\end{array}$
\caption{
Minimum central mass ($m_*$)
   for clusters with $F_2=3$
   and various outer radii.
Left-hand panel shows the effect of spatially uniform heating ($\nu=0$),
   and right-hand panel shows concentrated heating with $\nu=2$.
AGN power
   (annotated as in Fig.~\ref{figure.agn.frac} and \ref{figure.agn.mass})
   makes little difference to the limiting $m_*$.
The dashed curve shows results without a cD galaxy.
}
\label{fig.mass.F3}
\end{figure}

\begin{figure}
\centering
$\begin{array}{cccc}
\includegraphics[width=84mm]{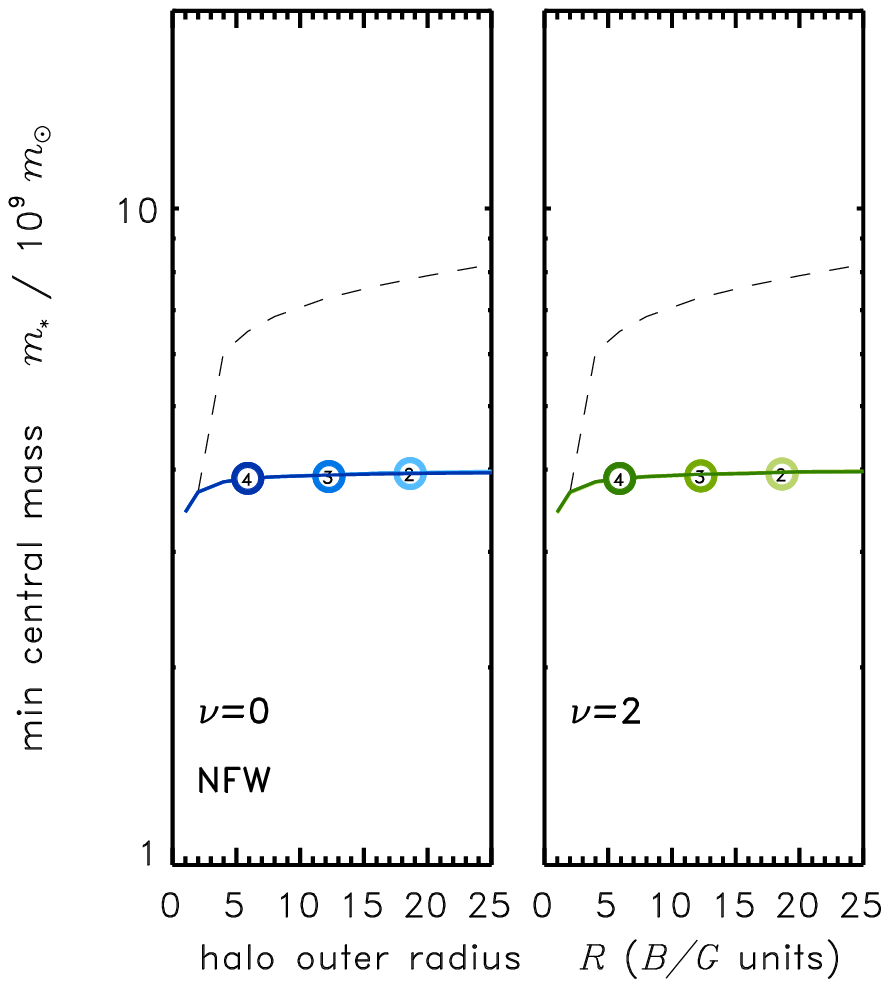}
\end{array}$
\caption{
Minimum central mass ($m_*$) versus cluster radius
   as in Fig.~\ref{fig.mass.F3}
   but with a NFW halo profile.
The $m_*$ show slightly more variation with $R$ and AGN power
   than the $F_2=3$ models,
   but less than for the preferred $F_2=8$ scenario.
}
\label{fig.mass.NFW}
\end{figure}

In the absence of a cD galaxy,
   all cosmologically reasonable baryon fractions imply a minimum
   $m_*\ga10^{10}m_\odot$ for $F_2=3$,
   and $m_*\ga8\times10^9m_\odot$ for NFW models
   (dashed lines in Figs~\ref{fig.mass.F3}
   and \ref{fig.mass.NFW} respectively).
The gravitational presence of the standard cD galaxy
   lowers these limits to
   $m_*\approx4\times10^9m_\odot$.
The cluster radius $R$ has negligible effect on these limits.
The power and functional form of the AGN heating
   has little further effect on $m_*$ limits
   for $F_2=3$ and NFW models:
   e.g. the $10^{-4}U_L, 10^{-3}U_L$ and $10^{-2}U_L$ power curves
   are almost inseparable in the panels of
   Figs~\ref{fig.mass.F3} and \ref{fig.mass.NFW}.
For the NFW cases, raising the AGN output from zero
    up to $10^{-2}U_L$ makes a few percent difference in $m_*$,
    though not enough to be visible on the scale of Fig.~\ref{fig.mass.NFW}.
For $F_2=3$ (Fig.~\ref{fig.mass.F3}) the difference is much smaller.
In summary, for $F_2=8$, $F_2=3$ and NFW scenarios alike,
   we find that the $m_*$ limit is influenced
   more by the cD galaxy potential (at kpc radii)
   than by any AGN heating,
   but the AGN effects are greater for $F_2=8$ than the alternative haloes.

\begin{figure}
\centering
$\begin{array}{cccc}
\includegraphics[width=84mm]{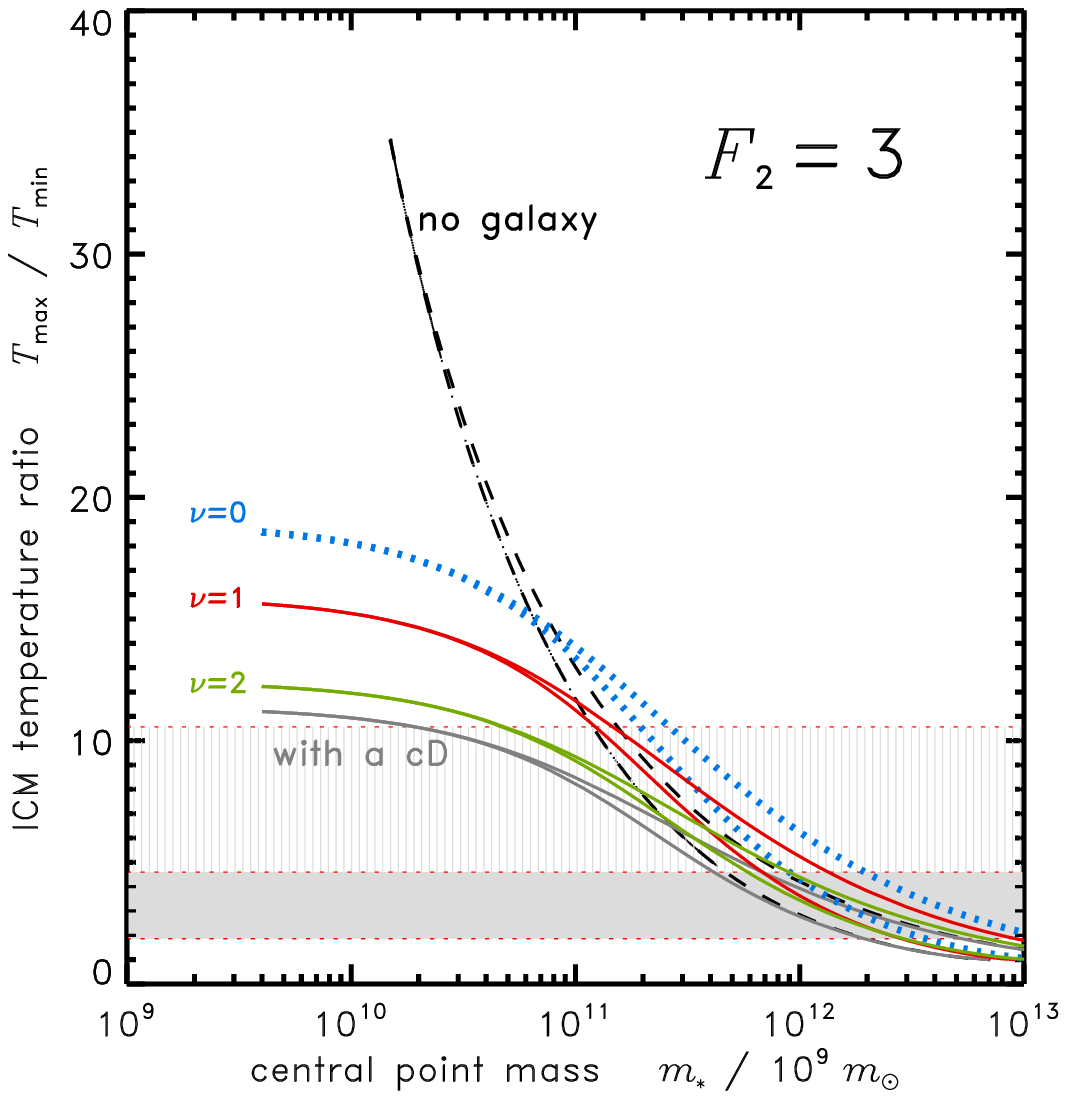}
\\
\end{array}$
\caption{
The ($T_\mathrm{max}/T_\mathrm{min}$) ratio of the ICM
   for cluster models with radius $R=16U_x\approx3.93$~Mpc
   and $F_2=3$ polytropic halo.
As in Fig.~\ref{fig.TmaxTmin},
   the grey shaded area is the range of ratios
   observed in cooling flow clusters,
   and the striped region includes the Centaurus~A group.
}
\label{fig.ToT.F3}
\end{figure}

\begin{figure}
\centering
$\begin{array}{cccc}
\includegraphics[width=84mm]{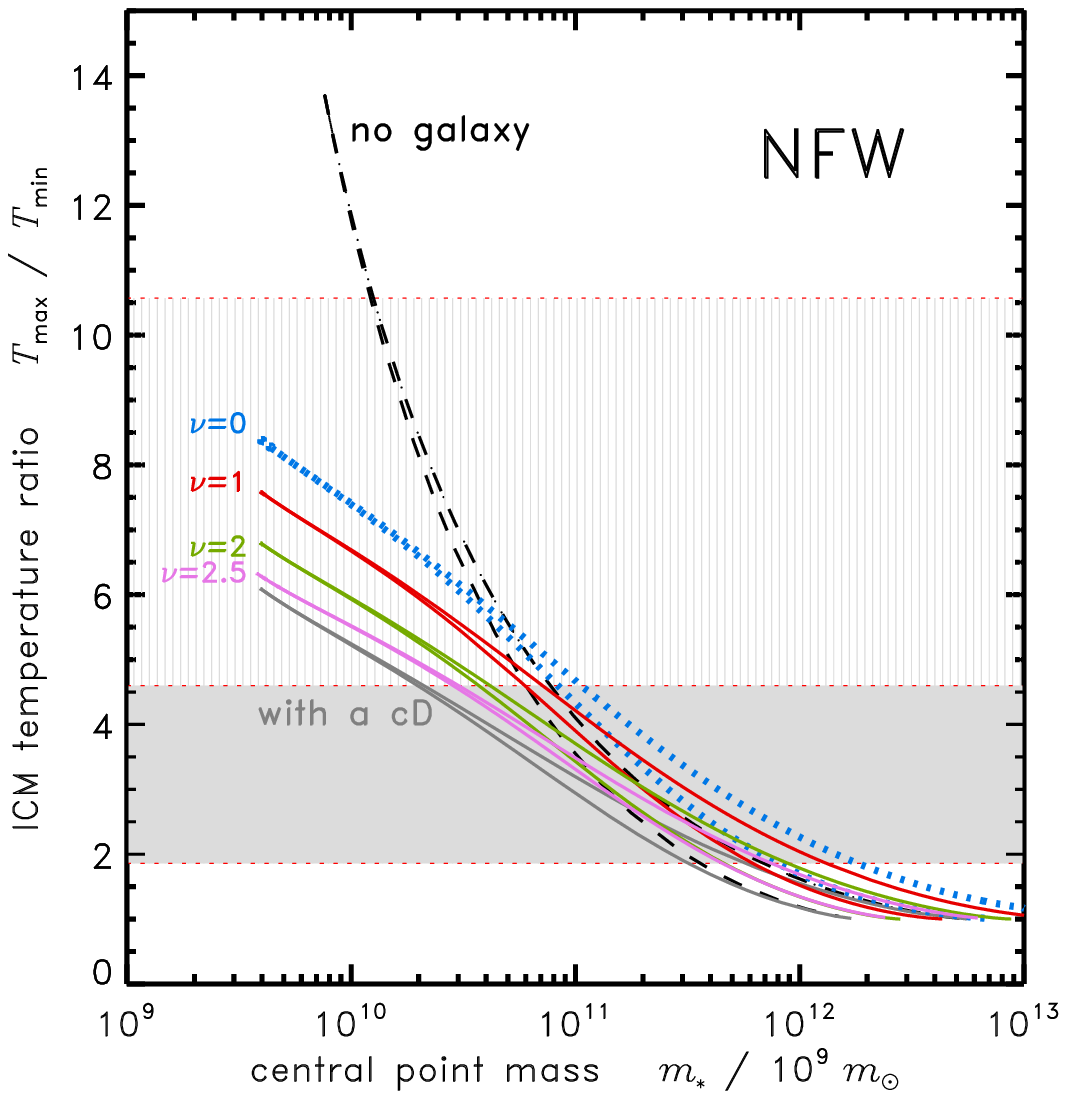}
\\
\end{array}$
\caption{
The ($T_\mathrm{max}/T_\mathrm{min}$) ratio of the intracluster gas 
   for cluster models equivalent to Fig.~\ref{fig.TmaxTmin}
   and Fig.~\ref{fig.ToT.F3},
   but with `NFW' dark halo profiles.
}
\label{fig.ToT.NFW}
\end{figure}

Fig.~\ref{fig.ToT.F3}
   is analogous to Fig.~\ref{fig.TmaxTmin}:
   showing the ICM temperature range $T_\mathrm{max}/T_\mathrm{min}$
   attainable in a set of self-consistent solutions of the $F_2=3$ clusters
   of outer radius $R=16U_x\approx3.93$~Mpc
   and different central masses above the minimum $m_*$.
Fig.~\ref{fig.ToT.NFW} shows the corresponding NFW models.
The curves depict cases without cD galaxy (dashed),
   with an inactive cD galaxy (grey),
   and various heating functions at high power
   ($10^{-2}U_L$, annotated by $\nu$).
Each scenario appears as two joined curves:
   they correspond to the `too fast' and `too cold' borders
   bounding the physically allowed zone
   in the configuration space.
At the minimum $m_*$ extreme,
   the greatest $T_\mathrm{max}/T_\mathrm{min}\la35$ for $F_2=3$,
   and $T_\mathrm{max}/T_\mathrm{min}\la14$ for NFW background.
We previously saw that AGN heating in $F_2=8$ clusters
   has little effect on the temperature ratio of the ICM.
Now in the $F_2=3$ and NFW models it appears that
   AGN heating {\em increases} $T_\mathrm{max}/T_\mathrm{min}$
   relative to the case of an inactive galaxy.
This effect is weaker for more concentrated heating (larger $\nu$).
The gas temperature range
   only agrees with observed values of large clusters
   ($2.5\la T_\mathrm{max}/T_\mathrm{min}\la4.5$, the grey shaded band)
   if the central point-mass
   is comparable to the combined mass of the stars,
   $m_*\ga M_\bigstar/2$.
To achieve a realistic $T_\mathrm{max}/T_\mathrm{min}$,
   the $F_2=8$ halo model only requires
   a central object of a few $10^8m_\odot$.
On these grounds, we re-emphasize that
   the large-$F_2$ regime
   is more plausible than NFW or $F_2=3$ models.
We would encourage astroparticle theorists
   to focus on candidate particles and fields
   that would naturally provide forms of dark matter
   with properties in the interval of $7\la F_2<10$.

\section{X-ray temperature fits}
\label{s.fitting}

The formulation presented in this work can also be used
   in parameter extractions from observational data,
   thus enabling comparisons between models and observations.
Here, we show a simple conceptual demonstration in which
   temperature profiles derived from X-ray observations
   are fitted by the cluster model discussed in this paper.
We extract projected gas temperature profiles
   (and error bars) from the data files
   in an online version of \cite{vikhlinin2006}.
For the purpose of this exercise,
   we will treat the two-dimensional projected temperature profile
   as representative of the three-dimensional spherical profile.

Assuming $F_2=8$ and our standard outer boundary conditions,
   we vary presence/absence of the cD galaxy,
   and the halo radius $R=4, 6, 8, 10, 12, 14, 16U_x$.
The baryon fraction is not controlled in these tests;
   we allow whatever values of $M_1$ and $M_2$
   happen to emerge from $m_*$-minimization.
We gather the cluster solutions at different $\Mcal_R$
   just along the acceptable side of the `too fast' border in configuration space.
For each model, we compute the $\chi^2$ best fit
   to the projected $T(r)$ data,
   allowing a linear auto-normalization process
   to obtain the best-fitting mass scale.
In effect, the fit has one continuous fitting variable
   ($\Mcal_R$, since optimization of $s_2$ is implied)
   and coarse trial values for $R$.

Despite this inflexibility,
   the reduced $\chi^2/\mathrm{d.o.f.}\approx2$.
We might expect tighter fits (and perhaps too much parameter degeneracy)
   if the baryon fraction, $\dot{m}$, $T_R$ and $R$ were all varied continually
   and independently of $\Mcal_R$ and the mass normalization.
A383 is the most massive of the three example systems fitted,
   with $R=8U_x\approx2.0$Mpc,
   and auto-normalization giving mass of
	$M=2.2\times10^{14}m_\odot$
   and inflow
	$\dot{m}=5.0~m_\odot~\mathrm{yr}^{-1}$
   ($\chi^2=21.9$ for 9 bins).
For A1991,
	$R=6U_x\approx1.5$Mpc,
	$M=8.2\times10^{13}~m_\odot$ and
	$\dot{m}=1.1~m_\odot~\mathrm{yr}^{-1}$
   ($\chi^2=25.3$ for 10 bins).
For USGC~S152,
	$R=4U_x\approx0.98$Mpc,
	$M=1.3\times10^{13}~m_\odot$ and
	$\dot{m}=0.068~m_\odot~\mathrm{yr}^{-1}$
	($\chi^2=15.5$ for 9 bins).
Note that the value of $\dot{m}^3/M^2$ is implicitly fixed constant.
Unlike the original semi-parametric modelling by \cite{vikhlinin2006},
   we do not need to omit any of the innermost points.
This inclusion enables our fits with a DM core rather than cusp.
The central rise in temperature
   (in some objects, e.g. in USGC~S152)
   emerges naturally in our formulation.
The uphill and downhill slopes around the temperature dip and peak
   also turn out to be similar to those appearing in nature.

\begin{figure}
\centering
$\begin{array}{cccc}
\includegraphics[width=84mm]{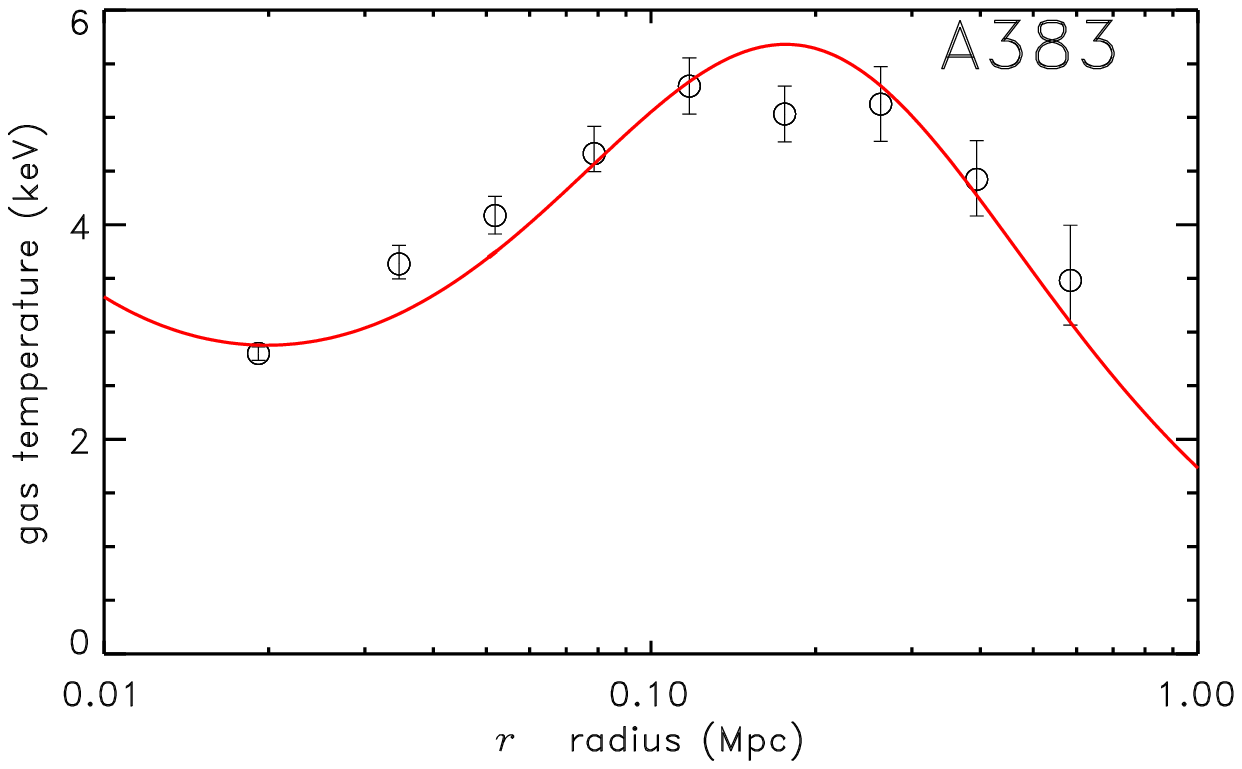}
\\
\includegraphics[width=84mm]{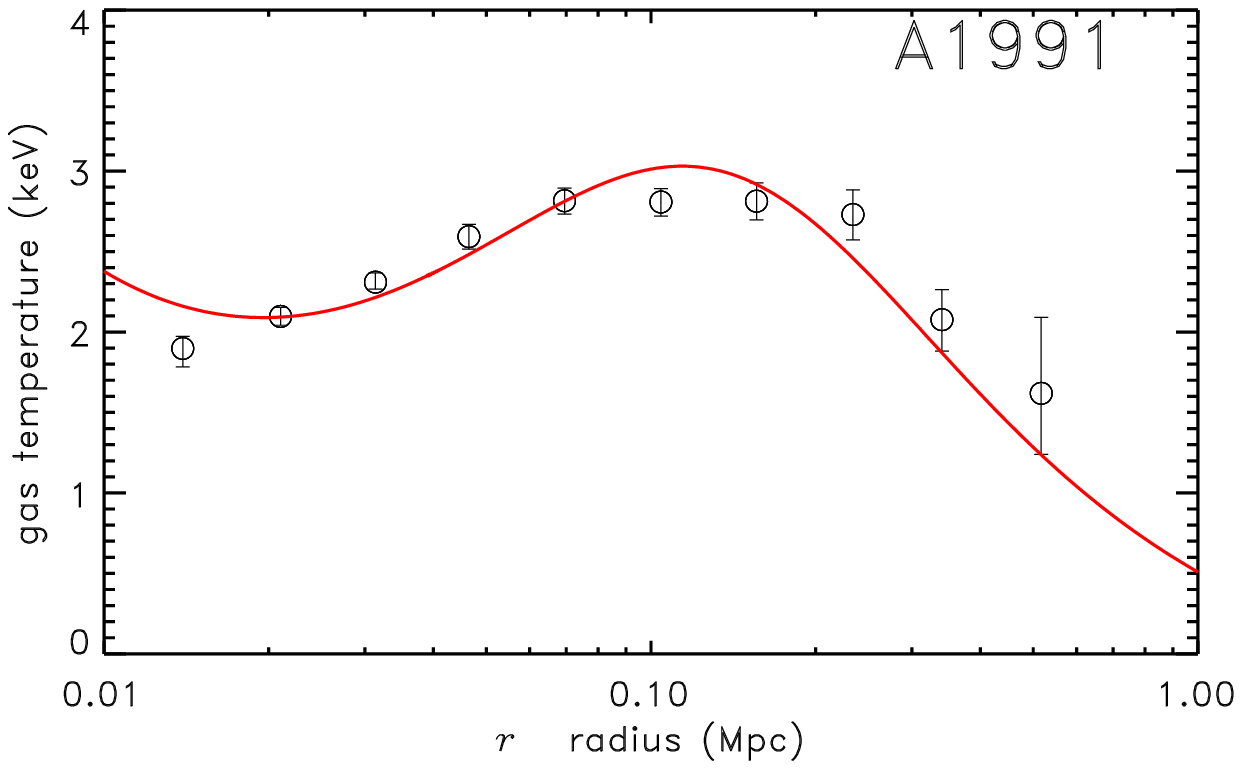}
\\
\includegraphics[width=84mm]{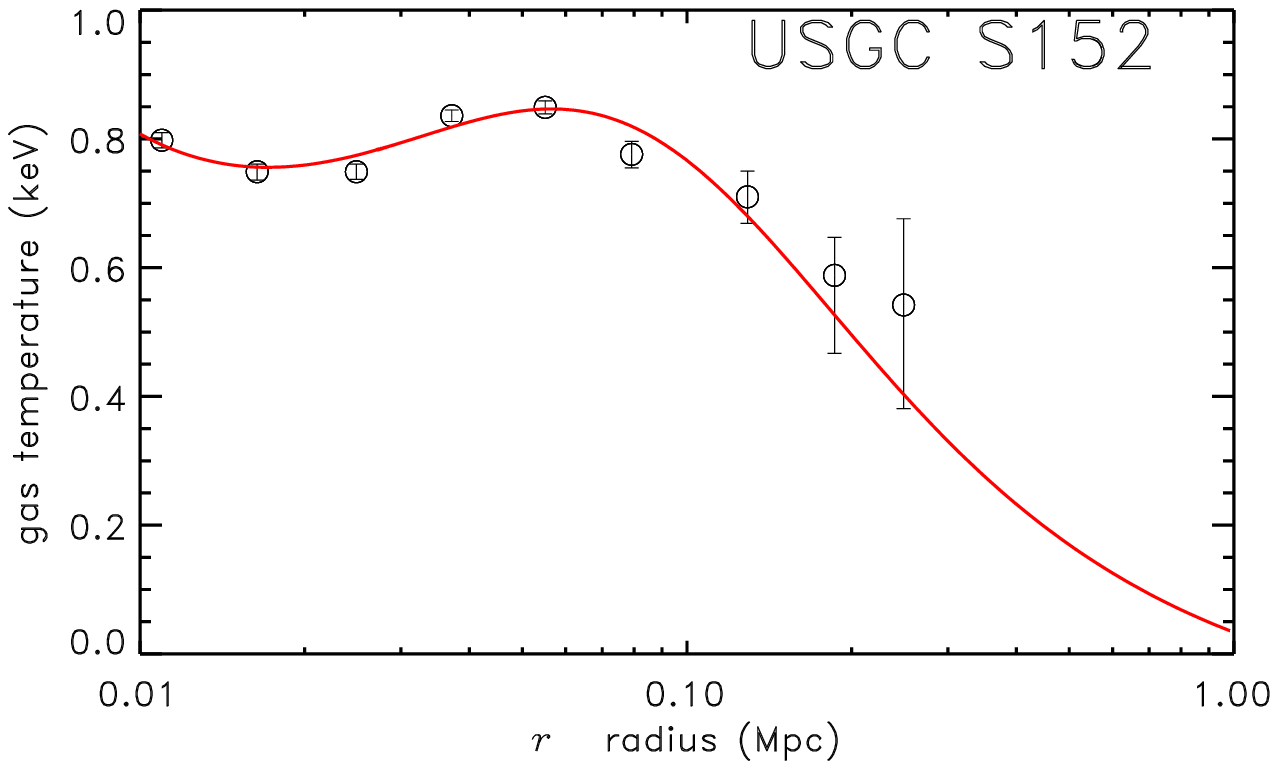}
\\
\end{array}$
\caption{
Cluster model fits to three of the cluster temperature profiles
   from Vikhlinin et~al. (2006).
The X-ray derived temperatures are projected,
   but the model profiles are three-dimensional.
}
\label{fig.V2006}
\end{figure}

\label{lastpage}
\end{document}